\documentclass[12pt]{article}

\usepackage{newtxtext,newtxmath}

\usepackage[T1]{fontenc}

\usepackage{graphicx}

\usepackage{makecell}
\usepackage{longtable}

\usepackage[letterpaper,margin=1in]{geometry}

\renewenvironment{abstract}
	{\quotation}
	{\endquotation}

\date{}

\makeatletter
\renewcommand{\fnum@figure}{\textbf{Figure \thefigure}}
\renewcommand{\fnum@table}{\textbf{Table \thetable}}
\makeatother

\usepackage{scicite}

\usepackage{url}

\def\scititle{
	Twisted-pair unilateral reconnection: A unifying driver for magnetically powered astrophysical bursts
}
\title{\bfseries \boldmath \scititle}

\author{
	Shu-Ping~Yan$^{1\ast}$,
	Wenhui~Yu$^{1,2}$,
	Jing~Ye$^{3,4\ast}$,
	Siming~Liu$^{5}$,
	Yang~Su$^{1,2}$,
	Huirong~Yan$^{6,7}$,\and
	Ping~Zhang$^{8}$,
	Guotianci~Xu$^{9,1}$,
	Jun~Lin$^{3,4}$,
	Haisheng~Ji$^{1,2}$,
	Zongjun~Ning$^{1,2}$,\and
	Wei~Chen$^{1,2}$,
	Li~Ji$^{1,2}$,
	Jingxing~Wang$^{3,4}$\and
	\small$^{1}$Purple Mountain Observatory, Chinese Academy of Sciences, Nanjing 210023, China.\and
	\small$^{2}$School of Astronomy and Space Science, University of Science and Technology of China, Hefei 230026, China.\and
	\small$^{3}$Yunnan Observatories, Chinese Academy of Sciences, Kunming 650216, China.\and	
	\small$^{4}$Yunnan Key Laboratory of Solar Physics and Space Science, Kunming, Yunnan 650216, China.\and	
	\small$^{5}$Southwest Jiaotong University, Chengdu 611756, China.\and	
	\small$^{6}$Deutsches Elektronen-Synchrotron DESY, Zeuthen, Germany.\and
	\small$^{7}$Institut für Physik \& Astronomie, Universität Potsdam, Potsdam, Germany.\and
	\small$^{8}$Wuhan University of science and technology, Wuhan 430081, China.\and
	\small$^{9}$School of Computer Science, Software Engineering and Cyberspace Security, Nanjing University of Posts\and\small and Telecommunications, Nanjing 210023, China.\and	
	\small$^\ast$Corresponding author. Email: yanshuping@pmo.ac.cn; yj@ynao.ac.cn\and
}

\begin{document} 

\maketitle

\begin{abstract} \bfseries \boldmath

Magnetic reconnection in twisted loops has long been invoked as an engine powering energetic transients from black hole accretion to neutron star mergers, yet never directly observed. Here we report the first direct observation of the complete reconnection of this type in a solar flare. We find a magnetic loop twisted to $\sim$540°, far exceeding the 180° twist assumed in existing simulations. This extreme twist inherently enables efficient multiple X‑line reconnection, akin to the role of turbulence in contemporary theory. Remarkably, the intertwined end breaks unilaterally after reconnection (unlike symmetric breaking in simulations), forming open field lines that release hot plasma---providing a promising mechanism for coronal generation or heating. We first detect hard X‑ray emission from the current sheet, directly proving it as a particle accelerator. Moreover, we discover a power‑law relationship between quasi‑periodic oscillation frequency and magnetic field strength across solar flares, black hole binaries, active galactic nuclei, magnetars, and gamma‑ray bursts. This relation identifies twisted-pair unilateral reconnection as a common burst mechanism and provides a natural ruler for cosmic magnetic fields. These findings establish an observational foundation for future reconnection theory and simulations, offering a unified framework for magnetically powered bursts.

\end{abstract}

\noindent

Magnetic reconnection powers a wide range of energetic explosions in the universe, from Earth's magnetospheric substorms to violent outbursts in distant systems~\cite{Zhang2011, Ji2011}. In particular, twisted magnetic loops serve as a fundamental reconnection geometry that drives high‑energy bursts across diverse astrophysical sources, including black hole binaries, active galactic nuclei, magnetars, and gamma‑ray bursts~\cite{Romanova1998, Donati2005, Matteo1999, Jacquemin2024, Thompson2002, Parfrey2013, Kaspi2017, Carrasco2019, Lyubarsky2020, Mahlmann2023, Most2022, Most2023}. However, these sources are too distant to observationally confirm whether such bursts are indeed caused by twisted‑loop reconnection.

Among all accessible environments, solar flares offer an ideal natural laboratory for studying magnetic reconnection, as they are close enough to allow fine structural resolution. The flare reconnection geometries observed on the Sun are complex and can be grouped into three categories: (i) reconnection driven by a destabilized, upward‑moving flux rope that lifts magnetic loops; (ii) reconnection between pre‑existing and newly emerging magnetic field lines; and (iii) internal reconnection within a sigmoidal, twisted flux rope~\cite{Lin2000,Priest2000,Shibata2011}. The twisted‑loop reconnection geometry described above has never been observed. Moreover, in none of these cases has the detailed process of how magnetic field lines reconnect ever been directly observed.

In the prevailing flare model, magnetic reconnection first generates a current sheet, which is then fragmented by the tearing‑mode instability into magnetic islands where particles are subsequently accelerated~\cite{Drake2006,Ji2022}. This picture is strongly supported by observations. Ultraviolet imaging has directly captured the current sheet (as a plasma sheet) and the magnetic islands (as plasmoids)~\cite{Lin2005,Su2013,Lu2022,Yan2022,Kumar2025}. Meanwhile, hard X‑ray (HXR) bremsstrahlung from the accelerated electrons has been detected from chromospheric footpoints, coronal loops, and faint coronal sources near the current sheet~\cite{Masuda1994,Sui2004,Aschwanden2005,Krucker2008,Liu2008,Holman2011,Su2013,Chen2020,Kumar2025}. However, despite the model predicting that particles are accelerated within the current sheet, no HXR emission has ever been directly detected from the current sheet.

Quasi-periodic pulsations (the solar analogue of astrophysical quasi‑periodic oscillations; hereafter both referred to as QPOs) are frequently observed in solar flare emissions and may probe current sheet instabilities~\cite{McLaughlin2018,Zimovets2021,Kou2022,Kumar2025}. Nevertheless, the mechanism responsible for QPOs in the current sheet remains poorly understood, primarily due to insufficient time and frequency resolution in the available measurements.

In summary, the twisted‑loop reconnection predicted to power many astrophysical bursts has never been directly observed, nor has the complete fine‑scale reconnection process of magnetic field lines. Observations confirm the formation of current sheet, magnetic islands, and particle acceleration in solar flares, yet direct evidence that the current sheet itself serves as a primary particle accelerator remains elusive, and the physical origin of QPOs from the current sheet remains unclear.

\subsection*{Twisted-pair unilateral reconnection}
\noindent

The solar flare SOL2015-08-27T05:45 occurred on the southwest solar disk (Fig.~\ref{fig:FlarePos}) and was observed by GOES, RHESSI, SDO/AIA, NoRP, and NoRH~\cite{methods}. Its GOES 1--8~\AA\ soft X‑ray flux peaked at M2.9 near 05:45~UT~\cite{Zhang2018}.

\begin{figure}[htbp]
    \centering
    \includegraphics[width=1.0\textwidth]{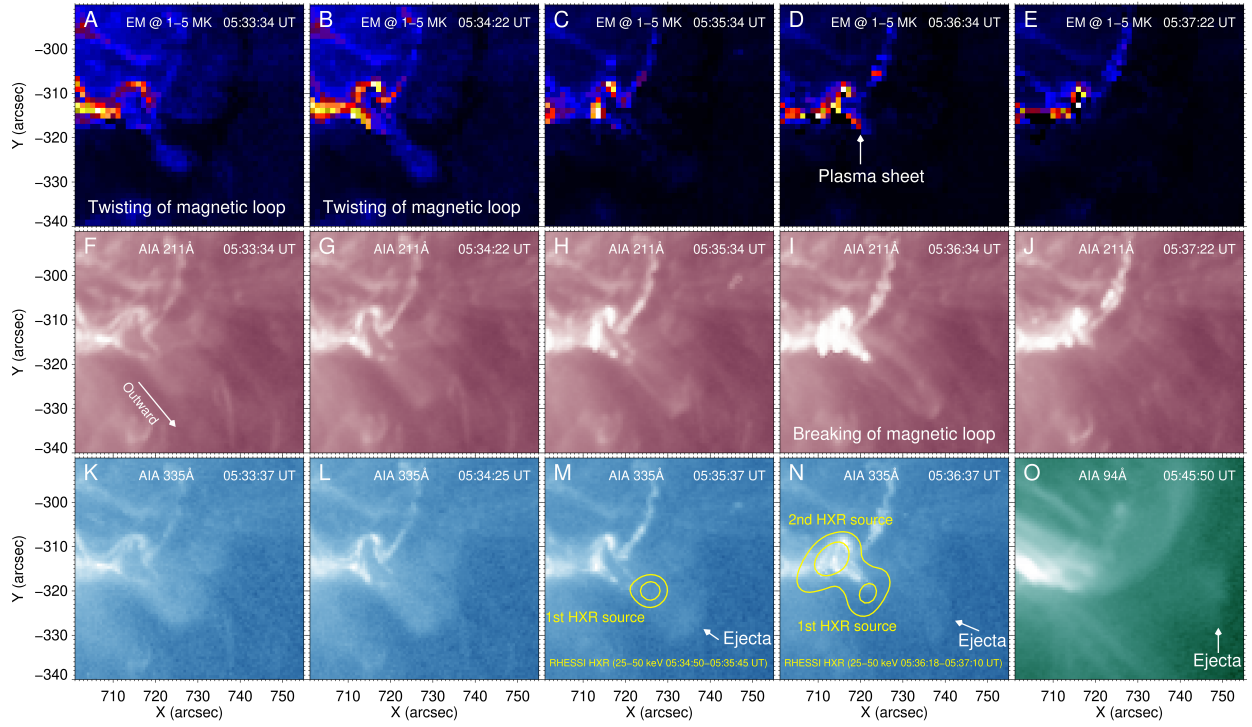}
    \caption{\textbf{The twisted-pair unilateral reconnection in solar flare SOL2015-08-27T05:45.}
    Each panel shows a snapshot of the flare's evolution, arranged from left to right in chronological order. The white arrow in (f) shows the direction outward from the sun's surface.
    (\textbf{A} to \textbf{E}) Total emission measure in the 1--5~MK temperature range. The emission measure shows a magnetic loop that twists to 540° and then reconnects, forming a plasma sheet. A plasma sheet is a current sheet composed of plasma.
    (\textbf{F} to \textbf{J}) AIA 211~\AA\ images displaying a magnetic loop that breaks unilaterally during reconnection, forming open field lines.
    (\textbf{K} to \textbf{O}) AIA 335~\AA\ and 94~\AA\ images with RHESSI 25--50~keV hard X-ray (HXR) contours overlaid. An HXR source forms in the current sheet region, followed by a second HXR source beneath the first. Outward ejecta form during reconnection (M and N). After the magnetic loop breaks unilaterally, the ejecta deflect significantly toward the break (O).}
    \label{fig:Flarevolution}
\end{figure}

To visualize the reconnection process, we examined SDO/AIA and RHESSI data. Movie S1 shows the evolution of the differential emission measure (DEM) for plasma temperatures between 1 and 5 MK during the main phase, as derived from AIA imaging; Movies~S2 to S10 are the corresponding single‑wavelength animations. Movies~S11 to S13 (pre‑flare) and S14 to S16 (late phase) illustrate the triggering magnetic activity and post‑flare structures. Figure~\ref{fig:Flarevolution} presents a time‑lapse portrait of the main phase.

From this analysis, the magnetic loop is twisted to $\sim$540° (Fig.~\ref{fig:Flarevolution}, A and B; movie~S1), far exceeding the $\sim$180° typically assumed in reconnection simulations~\cite{Parfrey2013, Mahlmann2023, Most2022, Most2023}. Such extreme twist stores more magnetic energy~\cite{Kaspi2017}, implying that weaker field strengths can achieve the same luminosity. 

The highly twisted magnetic loop ensures a broad reconnection region and multiple X‑lines, which directly enables fast reconnection. In contrast, current theories rely on turbulence to achieve a similarly broad reconnection region and numerous X‑lines, thereby ensuring high reconnection efficiency~\cite{Lazarian1999}. Of course, this does not preclude a role for turbulence in further enhancing the reconnection efficiency.

This twist drives a new reconnection pattern: the loop breaks unilaterally, producing open field lines instead of detached closed loops (Fig.~\ref{fig:Flarevolution}I; movie~S2). Outward ejecta tilt toward the break side after the reconnection (Fig.~\ref{fig:Flarevolution}, M to O; movie~S3). A plasma sheet appears primarily at 1--5~MK (Fig.~\ref{fig:Flarevolution}D; fig.~\ref{fig:AIA_EM}), indicating current sheet plasma heated to that range. Hot current sheet plasma, together with accelerated particles, can escape directly into the corona via open field lines~\cite{Judge2024}. This offers a potential explanation for the observed scarcity of upward‑escaping electrons (detected near Earth) relative to downward‑precipitating HXR‑producing electrons~\cite{Krucker2007}.

\begin{figure}[htbp]
    \centering
    \includegraphics[width=1.0\textwidth]{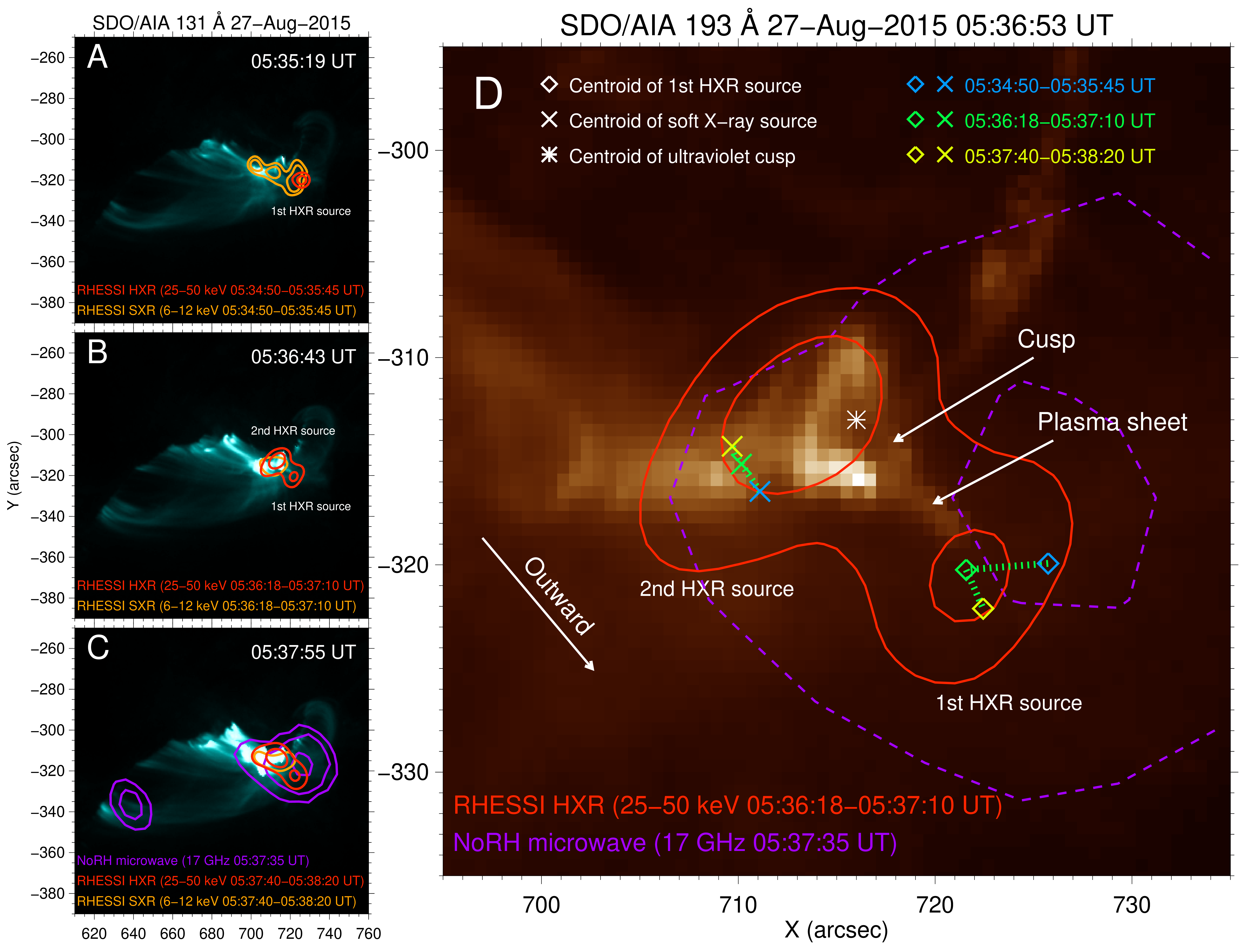}
    \caption{\textbf{X-ray and microwave emissions from the flare.}
    (\textbf{A}) X-ray sources during the flare's precursor phase. Red contours (50\% and 80\% of the maximum) show HXR (25--50~keV) emission. Gold contours (60\% and 80\% of the maximum) show 6--12~keV soft X-ray emission.
    (\textbf{B}) X-ray sources during the early peak of the HXR emission. HXR emission is shown as red contours (30\% and 60\% of the maximum). Soft X-ray emission is shown as gold contours (45\% of the maximum). The second HXR source emerges.
    (\textbf{C}) X-ray and microwave emissions during the late peak of the HXR emission. HXR emission is shown as red contours (25\% and 55\% of the maximum). Soft X-ray emission is shown as gold contours (55\% of the maximum). Purple contours (30\%, 50\%, and 90\% of the maximum) show 17~GHz microwave emission.
    (\textbf{D}) Spatial relationship between high-energy emissions and ultraviolet features. A plasma sheet, the manifestation of a current sheet, is visible in AIA 193~\AA, with a V-shaped cusp structure beneath it. Central positions are marked by diamonds (first HXR source), crosses (soft X-ray source), and asterisks (ultraviolet cusp). Symbol colors indicate the time intervals.}
    \label{fig:CSHXR}
\end{figure}

To further test whether HXRs originate from the current sheet, we selected three time intervals corresponding to distinct phases of the HXR burst, based on RHESSI data quality: 05:34:50–05:35:45~UT (precursor), 05:36:18–05:37:10~UT (early peak), and 05:37:40–05:38:20~UT (late peak). A weak HXR source appears in the twisting region during the precursor (Fig.~\ref{fig:Flarevolution}M; Fig.~\ref{fig:CSHXR}A) and brightens as reconnection proceeds (Fig.~\ref{fig:Flarevolution}N; Fig.~\ref{fig:CSHXR}, B or D). Beneath the current sheet, a cusp-shaped magnetic structure emerges across all AIA wavelengths (Fig.~\ref{fig:CSHXR}D; fig.~\ref{fig:CScusp}), consistent with reconnection theory. The HXR source lies above the cusp, co‑located with the current sheet. A second, more luminous HXR source appears at a lower altitude, suggesting that some electrons accelerated within the current sheet produce the first source, while others are injected into the denser plasma below to generate the second.

The first HXR source centroid is $8.1\pm0.7$~Mm from the cusp centroid and $\sim$4~Mm above the cusp peak (Fig.~\ref{fig:CSHXR}D; table~\ref{tab:centroids}). Its distribution is confined along the lower half of the current sheet (fig.~\ref{fig:CSevolution}, D and E), ruling out an origin above the current sheet. Distances to the soft X‑ray source are $13.4\pm1.9$, $11.1\pm0.8$, and $13.3\pm1.8$~Mm for the three intervals---much smaller than the typical loop‑top to high coronal HXR source separation~\cite{Sui2004,Su2013}. This confirms that the first HXR source originates from within the current sheet.

The reconnecting twisted loop internally connects to a heart‑shaped magnetic structure on a flux rope, which passes through underlying parallel loops (fig.~\ref{fig:CSevolution}; movies~S2 to S8; schematic in fig.~\ref{fig:Flarepattern}). This directly observes a flare driven by reconnection above a flux rope, extending previous detections of energy release above a filament~\cite{Ji2003}. The heart‑shaped structure emerges about 35~minutes before the flare, likely from reconnection of intertwined loops, accompanied by weak ultraviolet bursts (fig.~\ref{fig:LC211}; movies~S11 to S13), suggesting that such ultraviolet eruptions may act as precursors. Late‑phase ultraviolet eruptions (movies~S14 to S16) indicate continued reconnection; both the heart‑shaped structure and the eruption site are co‑located at the intersection of the flux rope and a large loop.

The RHESSI spectra are remarkably soft and fit by either a thermal‑plus‑nonthermal or a two‑thermal model (fig.~\ref{fig:RHESSIspec}). Both include a $\sim$20~MK thermal component. In the two‑thermal model, the second component has a temperature of $\sim$52~MK in the early peak. If the first HXR source were part of this hot thermal component, the current sheet plasma would produce significant 9--15~MK emission, but such emission is very weak (fig.~\ref{fig:AIA_EM}D). Hence, the first HXR source is non‑thermal, with a fitted spectral index $\delta \approx$8.8, consistent with soft spectra ($\delta \gtrsim$4) associated with reconnection~\cite{Oka2018}.

Two distinct 17~GHz sources (purple contours in Fig.~\ref{fig:CSHXR}C) are evident. The right source resides at the same altitude as the first HXR source (Fig.~\ref{fig:CSHXR}D), indicating a shared current sheet origin. Analysis reveals a temperature of $\sim$1.7~MK, a non‑thermal electron spectral index of $\sim$5, and a magnetic field strength $<$60~G (fig.~\ref{fig:NoRH}). The left 17~GHz source shows $\sim$1~MK, a similar index ($\sim$5), but a stronger field of $\sim$200--600~G.

\subsection*{QPOs from reconnection current sheet}
\noindent

To further investigate flare current sheet dynamics, we applied the parameter-optimized Hilbert--Huang transform (PHHT)~\cite{Yan2024} and wavelet analysis to HXR, microwave, and ultraviolet light curves. While the traditional Hilbert--Huang transform is designed for gradually varying signals and depends on existing knowledge or experiments to ensure validity, the extended PHHT robustly extracts weak signals from explosive, large-amplitude variations without empirical validation.

Both methods independently detected millihertz (mHz) QPOs. Wavelet time‑frequency spectra show a pronounced enhancement at $\sim$5--7 mHz (figs.~\ref{fig:QPPwpp}B and \ref{fig:UVtf}B), which is consistently present in the PHHT time‑frequency spectra (figs.~\ref{fig:QPPwpp}C and \ref{fig:UVtf}C). The QPOs appear as narrow peaks in the PHHT marginal spectra, with the first intrinsic mode function (noise‑dominated) excluded (fig.~\ref{fig:QPPwpp}D). Their centroid frequencies are $5.14\pm0.01$ mHz (25--50 keV), $6.26\pm0.03$ mHz (2+3.75+9.4 GHz), and $7.37\pm0.01$ mHz (171+211+304+335+1600~\AA) (fig.~\ref{fig:QPPfre}). The consistent detection across four instruments and two methods confirms robust $\sim$5--7 mHz QPOs.

\begin{figure}[htbp]
  \centering 
  \includegraphics[width=0.8\textwidth]{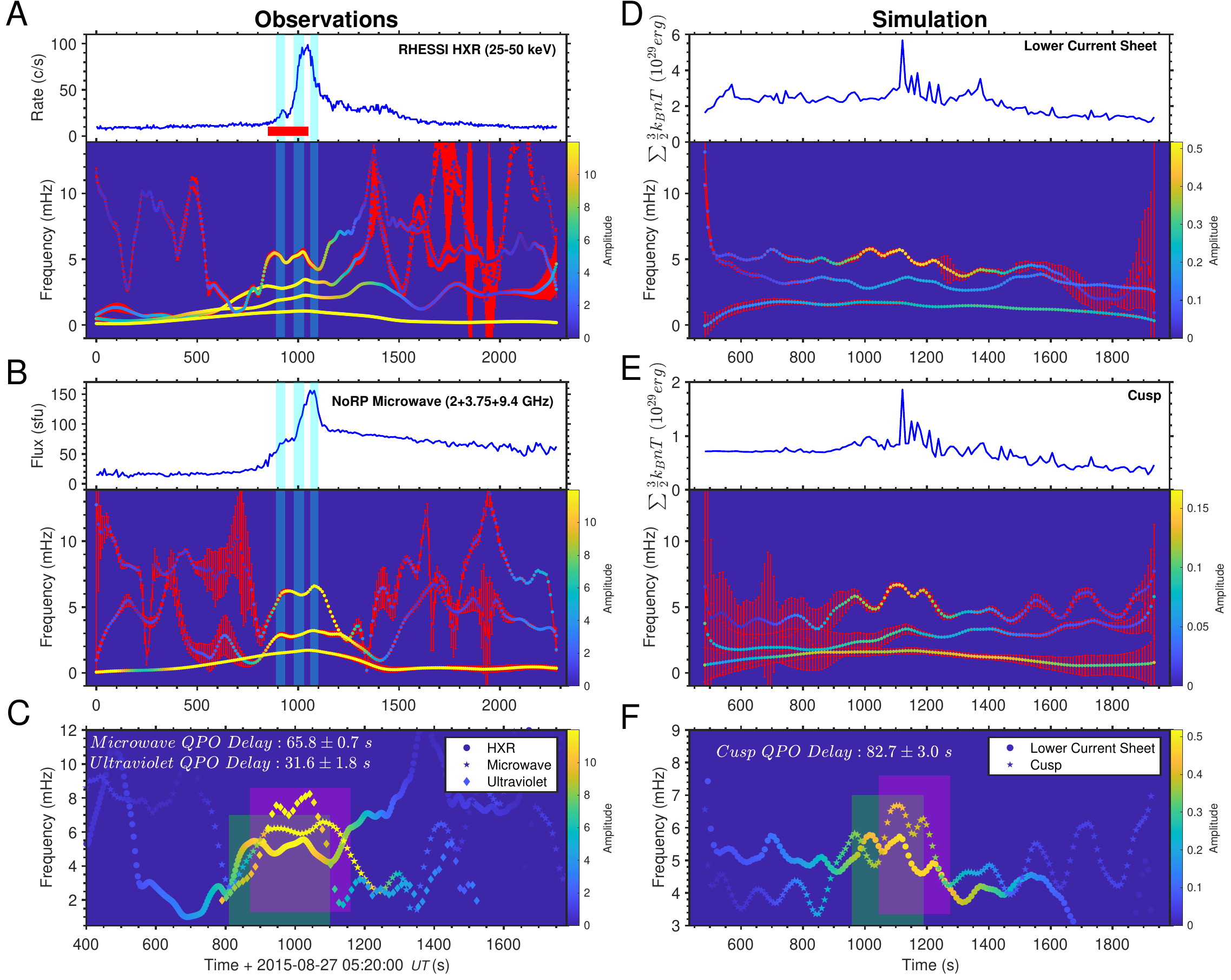}
    \caption{\textbf{Temporal delays of QPOs in observations and simulation.}
    (\textbf{A}) HXR light curve and its parameter-optimized Hilbert–Huang transform (PHHT) spectrum. The colored curves represent averaged PHHT spectra, with red error bars indicating standard deviations. For prominence, intrinsic mode functions preceding those containing the $\sim$5--7 mHz QPO are not shown in all PHHT spectra. Cyan bars mark the three time intervals used for HXR imaging analysis. The red horizontal bar indicates the time interval during which the QPO amplitude is strong.
    (\textbf{B}) Microwave light curve and its PHHT spectrum.
    (\textbf{C}) Time lag between the QPOs. The green bar indicates the time ranges used for calculating the time lag between the HXR and ultraviolet QPOs. The magenta bar indicates the time range used for the microwave QPO time-lag calculation. The microwave and ultraviolet spectral amplitudes are scaled for clarity.
    (\textbf{D}) Temporal evolution of thermal energy in the lower current sheet from a flare magnetohydrodynamic simulation, with corresponding PHHT spectrum. The thermal energy is defined as $\sum \frac{3}{2}k_B nT$, where $k_B$ is Boltzmann constant, $n$ is the plasma number density, and $T$ is temperature. 
    (\textbf{E}) Thermal energy evolution in the cusp region and its PHHT spectrum.
    (\textbf{F}) Time lag between QPOs from the simulation. The green and magenta bars mark the time ranges of the QPO spectra from the lower current sheet and the cusp, respectively, used to calculate the time lag.}
    \label{fig:QPPdatasim}
\end{figure}

The HXR, microwave, and ultraviolet QPOs commence before the main HXR rise and show M-shaped time-frequency profiles (Fig.~\ref{fig:QPPdatasim}, A and B; fig.~\ref{fig:UVtf}C). Monte Carlo cross-correlation reveals that microwave QPO lag HXR QPO by $65.8\pm0.7$~s, and ultraviolet QPO lag by $31.6\pm1.8$ s (Fig.~\ref{fig:QPPdatasim}C). In contrast, the underlying light curves show different delays (microwave lags by $\sim$20~s, ultraviolet leads by $\sim$76~s; figs.~\ref{fig:LCcorrelate} and \ref{fig:lags2eng}), indicating that QPO time lags are distinct from bulk emission delays.

\begin{figure}[htbp]
    \centering
    \includegraphics[width=1.0\textwidth]{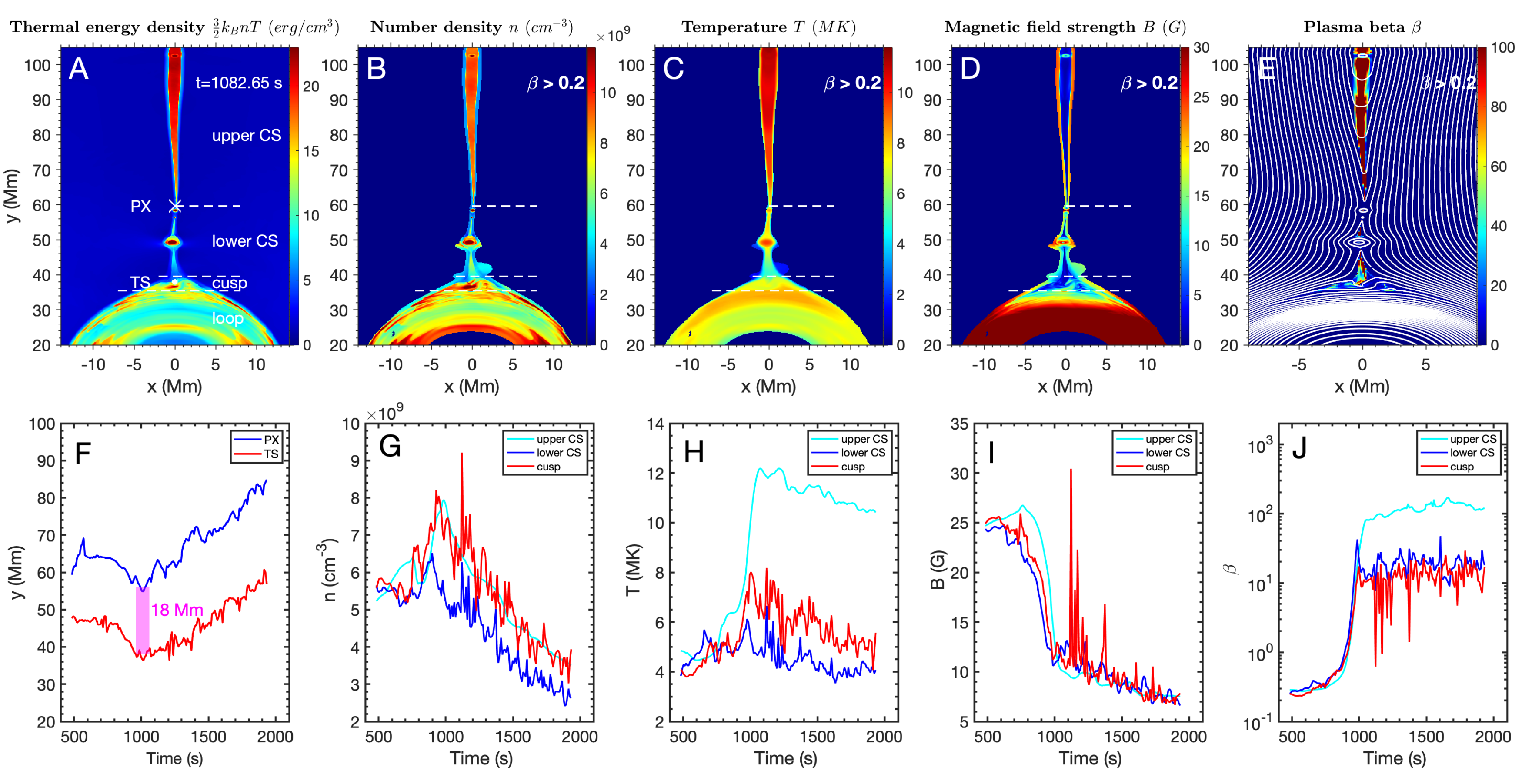}
    \caption{\textbf{Magnetohydrodynamic simulation of flare reconnection and its parameters.} 
	(\textbf{A}) Spatial distribution of the thermal energy at a representative time. The flare volume is divided into four regions based on the height of the principal X-point ($PX$) and the central position of termination shock ($TS$).
	(\textbf{B} to \textbf{E}) Spatial distributions in regions where plasma parameter $\beta >$0.2 at the representative time: number density (B), temperature (C), magnetic field strength (D), magnetic field lines and plasma $\beta$ (E).
	(\textbf{F}) Temporal evolution of $PX$ and $TS$ heights. The magenta bar indicates the separation between $PX$ and $TS$ during the early flare phase.
	(\textbf{G} to \textbf{J}) Temporal evolution of mean parameters in the upper current sheet, lower current sheet, and cusp (where $\beta >$0.2): number density (G), temperature (H), magnetic field strength (I), and plasma $\beta$ (J).}
    \label{fig:Simulpara}
\end{figure}

To investigate QPO origins, we performed a flare magnetohydrodynamic simulation (movie~S17; \cite{Ye2020}) and analyzed it with PHHT and wavelet. The simulated volume was divided into four regions based on the centroid positions of the principal X-point ($PX$) and termination shock ($TS$): loop (0 to $TS-2.5$~Mm), cusp ($TS-2.5$~Mm to $TS+1.6$~Mm), lower current sheet ($TS+1.6$~Mm to $PX$), and upper current sheet ($PX$ to 200~Mm) (Fig.~\ref{fig:Simulpara}A). The QPOs in thermal energy ($\propto$ plasma pressure) within the lower current sheet and cusp most closely match the observed HXR and microwave QPOs, respectively (Fig.~\ref{fig:QPPdatasim}, D and E). The simulated lag of the cusp QPO behind the lower current sheet QPO ($82.7\pm3.0$~s; Fig.~\ref{fig:QPPdatasim}F) agrees with the observed microwave--HXR delay ($\sim66$~s; Fig.~\ref{fig:QPPdatasim}C).

These results account for the observed delays relative to the HXR QPO: the ultraviolet delay ($\sim$32~s) is about half the microwave delay ($\sim$66~s), consistent with HXR, microwave, and ultraviolet QPOs originating from the current sheet, cusp, and a composite of both, respectively. No significant lag between 17~GHz and HXR QPOs (fig.~\ref{fig:QPP17G2hsi}) matches their co-spatial imaging (Fig.~\ref{fig:CSHXR}D).

Our analysis reveals a stronger association of the HXR and microwave QPO properties with thermal plasma parameters than with magnetic or kinetic ones (Fig.~\ref{fig:QPPdatasim}; fig.~\ref{fig:SimulEngwpp_withbkg}). Given that microwave emission arises from gyrosynchrotron radiation of energetic electrons, the tight correlation of these non-thermal QPOs with thermal parameters implies that the thermal plasma serves as the source of these electrons---that is, part of it is accelerated during reconnection. Thus, thermal dynamics are imprinted on both thermal (UV) and non‑thermal (HXR, microwave) radiation.

Closed magnetic structures (magnetic islands in 2D, loops in 3D) form within the current sheet (Fig.~\ref{fig:Simulpara}E) and propagate from the lower current sheet to the cusp in $\sim$40~s (movie~S18). Early in the flare, the $PX$ and $TS$ are separated by $\sim$18~Mm, with the $TS$ located $\sim$9~Mm below the centroid of the lower current sheet (Fig.~\ref{fig:Simulpara}F), matching the $\sim$8.1~Mm separation between the cusp centroid and current sheet HXR source centroid (Fig.~\ref{fig:CSHXR}D; table~\ref{tab:centroids}). Number density rises then declines in all regions, with upper current sheet and cusp values comparable and slightly higher than in the lower current sheet (Fig.~\ref{fig:Simulpara}G).

Cross-correlation shows that the density increase in the cusp lags that in the lower current sheet by $\sim$42~s, while the corresponding lag for temperature rise is shorter ($\sim$27~s; fig.~\ref{fig:Simulags}, A and B). Within each region, heating lags density increase by $\sim$81~s in the lower current sheet and $\sim$63~s in the cusp (fig.~\ref{fig:Simulags}, C and D). We interpret these lags as follows: (i) The 81~s heating delay in the lower current sheet reflects the timescale of magnetic‑to‑thermal energy conversion via dissipation or compression \cite{Priest2000}. (ii) The shorter $\sim$63~s lag in the cusp indicates more efficient thermal conduction across termination shocks. (iii) The 42~s cross‑region density lag matches the $\sim$40~s travel time of magnetic islands from the reconnection site to the cusp (movie~S18)).

The 83-second lag of the plasma pressure ($\propto nT$) QPO in the cusp relative to the lower current sheet exceeds both density ($\sim$42~s) and temperature ($\sim$27~s) delays, indicating that the oscillatory pressure signal is not transported solely by bulk flow or thermal fronts but emerges from nonlinear coupling of density and temperature perturbations.

Temperature and plasma $\beta$ are significantly elevated in the upper current sheet compared to the lower current sheet and cusp (Fig.~\ref{fig:Simulpara}, H and J), which facilitates particle escape. Magnetic field strength shows a steep initial drop followed by gradual decay across all regions, indicating magnetic energy consumption during the flare~\cite{Fleishman2020}.

\subsection*{A universal QPO--B power law}
\noindent

In this study, we report the first discovery of magnetic reconnection in twisted loops on the Sun. Theoretically, twisted-loop reconnection is central to outbursts from black hole X-ray binaries, active galactic nuclei, magnetars, and gamma-ray bursts~\cite{Romanova1998, Matteo1999, Jacquemin2024, Thompson2002, Parfrey2013, Kaspi2017, Carrasco2019, Lyubarsky2020, Mahlmann2023, Most2022, Most2023}. Yet direct confirmation from these distant sources remains impossible.



\begin{figure}[htbp]
    \centering
    \includegraphics[width=1.0\textwidth]{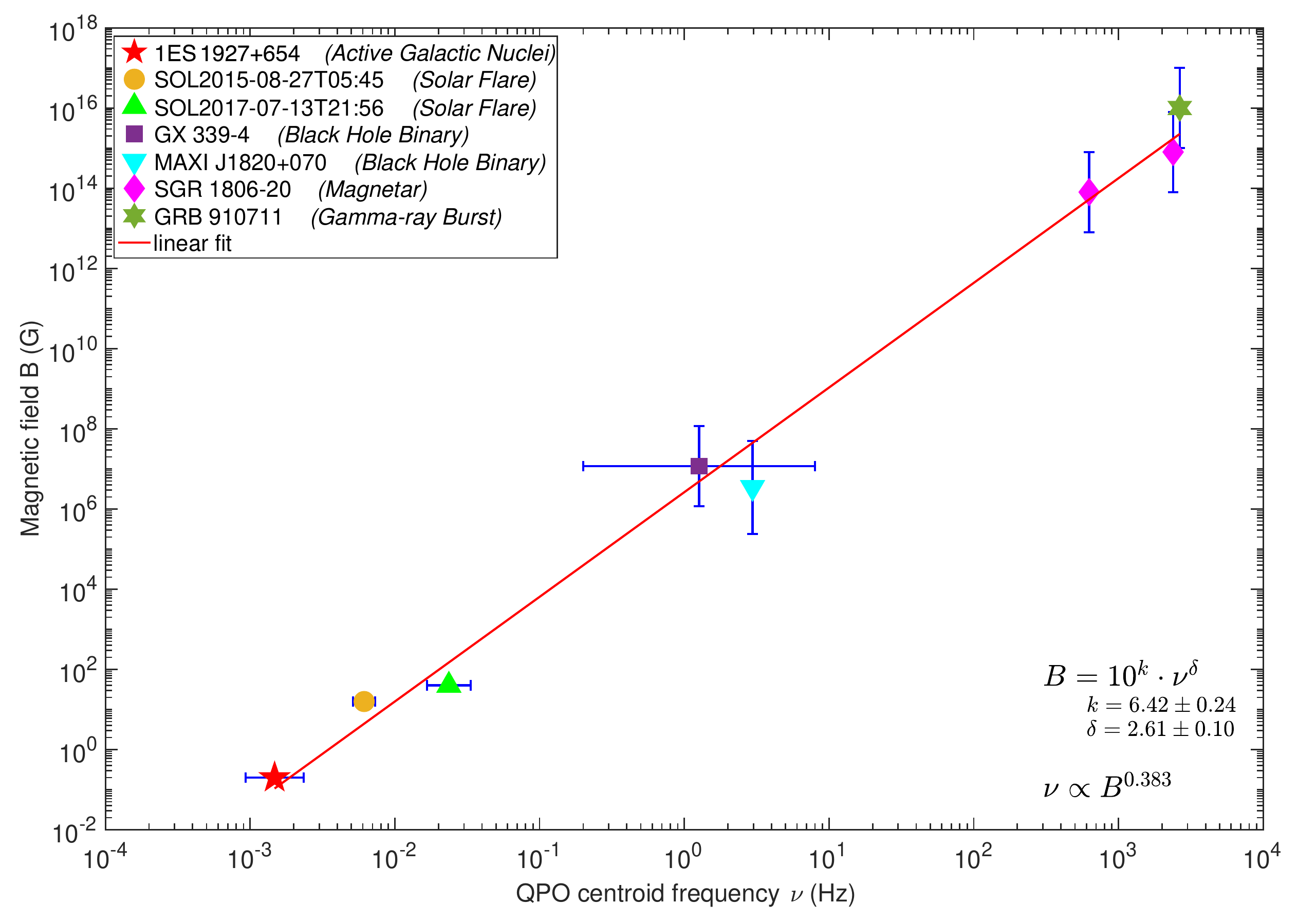}
    \caption{\textbf{Relationship between QPO frequency and magnetic field strength.} 
	Shown are all currently known pairs of QPO frequency and magnetic field strength. On a log--log scale, the relationship follows a linear trend (red line). A power-law fit $B = 10^k \cdot \nu^{\delta}$ yields $\delta = 2.61 \pm 0.10$ and $k = 6.42 \pm 0.24$ (adj. $R^2 = 0.99$, $p = 1.9 \times 10^{-7}$). The inverse relation $\nu \propto B^{1/\delta} = B^{0.383}$ gives an exponent of $0.383$, which is close to the golden ratio ($\phi \approx 0.382$). }
    \label{fig:Bfre}
\end{figure}

To explore this possibility, we compiled QPO frequencies and magnetic field strengths from the literature for a wide range of astrophysical sources (table~\ref{QBsource0}), identifying pairs where a QPO is associated with a specific field strength (table~\ref{QBsource1}; supplementary text). Plotting these QPO frequencies against the corresponding magnetic field strengths on a logarithmic scale (Fig.~\ref{fig:Bfre}) reveals a remarkable power-law relationship spanning more than 18 orders of magnitude in magnetic field strength. This relation holds across fundamentally different classes of objects---including the active galactic nucleus 1ES~1927+654, solar flare SOL2015-08-27T05:45, black hole X-ray binaries GX~339-4 and MAXI~J1820+070, magnetar SGR~1806-20, and gamma-ray burst GRB~910711.

The power-law fit yields $B = 10^{k} \nu^{\delta}$, with $k = 6.42 \pm 0.24$ and $\delta = 2.61 \pm 0.10$, where $B$ is in Gauss and $\nu$ in Hz. Despite vast differences in physical scales and burst energies, all these sources fall along the same trend. Although the magnetic field estimates carry uncertainties, they are grounded in reasonable physical assumptions. Crucially, the 18-order dynamic range renders the moderate uncertainties in individual estimates negligible for the overall scaling. The consistency across independently studied systems thus confirms that the relationship is physically meaningful.

Notably, the relationship can be expressed as $\nu \propto B^{0.383}$, where the exponent $0.383$ is remarkably close to the golden ratio ($\phi \approx0.382$). This unexpected concordance suggests a profound underlying order across cosmic scales.

The observed scaling suggests that magnetic reconnection universally governs QPO generation across these systems. In this process, magnetic energy is converted into thermal energy, heating the plasma, while the tearing-mode instability drives oscillations in plasma temperature and density. On the other hand, the thermal plasma supplies particles for acceleration. This complex interplay---combined with variations in plasma density, current sheet length, and magnetic resistivity---yields the observed power-law.


In summary, the power-law relationship between QPO frequency and magnetic field strength established here provides the first direct observational evidence linking twisted-loop magnetic reconnection to the outburst mechanisms of black hole X-ray binaries, active galactic nuclei, magnetars, and gamma-ray bursts. This result unifies disparate astrophysical phenomena under a common physical framework and opens new avenues for probing extreme plasma processes across the universe.



\clearpage

\clearpage 

%
\bibliography{QPO2Mag} 
\bibliographystyle{sciencemag}

%
%
%
%
%
%


\section*{Acknowledgments}
We acknowledge the use of publicly available data from NASA’s Goddard Space Flight Center and NAOJ's Nobeyama Solar Radio Observatory. We also gratefully acknowledge the developers of the following software: the empirical mode decomposition and CEEMDAN codes by G. Rilling, P. Flandrin, and M.A. Colominas, and the wavelet analysis code by C. Torrence and G. Compo. We thank Yalan Chen, Xiayu Zhang, Lei Lu, Zhentong Li, Shuinai Zhang, and Yan Li for helpful discussion. 
\paragraph*{Funding:}
S.-P.Y. acknowledges support from the National Natural Science Foundation of China (NSFC) under Grant U1838112. J.Y. acknowledges support by Strategic Priority Research Program of the Chinese Academy of Sciences No.XDB0560000, National Key R\&D Program of China No.2022YFF0503804, and NSFC grant 12573062.
\paragraph*{Author contributions:}
S.-P.Y. led the project, made and analyzed the AIA images and DEM movies, performed the PHHT and wavelet analyses, discovered the QPO--B relationship, and wrote the manuscript. J.Y. performed the magnetohydrodynamic simulation. W.Y. processed the AIA, RHESSI, and HMI image data, and performed the AIA DEM and RHESSI spectroscopic analyses. P.Z. processed the RHESSI and NoRP/NoRH light curves and their wavelet spectra. G.X. processed the NoRH image data and performed the corresponding spectroscopic analysis. S.L., Y.S., H.Y., J.L., H.J., Z.N., W.C., L.J., and J.W. offered valuable advices. All authors discussed the results and made contributions on manuscript writing.
\paragraph*{Competing interests:}
There are no competing interests to declare.
\paragraph*{Data and materials availability:}
The data used in this study are publicly available from the following sources:
\textit{SDO}/AIA and \textit{SDO}/HMI: \url{https://sdo.gsfc.nasa.gov/}.
\textit{RHESSI}: \url{https://hesperia.gsfc.nasa.gov/rhessi3/}.
\textit{NoRP}: \url{https://solar.nro.nao.ac.jp/norp/}.
\textit{NoRH}: \url{https://solar.nro.nao.ac.jp/norh/archive.html}.
The following software and codes were used for data processing and analysis:
\textit{SDO}: \url{https://www.lmsal.com/sdodocs/doc/dcur/SDOD0060.zip/zip/entry/}.
\textit{RHESSI}: \url{https://hesperia.gsfc.nasa.gov/rhessi3/}.
\textit{NoRP}: \url{https://solar.nro.nao.ac.jp/norp/}.
\textit{NoRH}: \url{https://solar.nro.nao.ac.jp/norh/}.
PHHT (modified from): \url{http://perso.ens-lyon.fr/patrick.flandrin/emd.html}.
Wavelet Transform: \url{http://paos.colorado.edu/research/wavelets/}.
\paragraph*{ License information:}


\subsection*{Supplementary materials}
Materials and Methods\\
Supplementary Text\\
Figs. S1 to S16\\
Tables S1 to S4\\
References \textit{(45-\arabic{enumiv})}\\ 
Movies S1 to S18\\


\newpage


\renewcommand{\thefigure}{S\arabic{figure}}
\renewcommand{\thetable}{S\arabic{table}}
\renewcommand{\theequation}{S\arabic{equation}}
\renewcommand{\thepage}{S\arabic{page}}
\setcounter{figure}{0}
\setcounter{table}{0}
\setcounter{equation}{0}
\setcounter{page}{1} 


\begin{center}
\section*{Supplementary Materials for\\ \scititle}

Shu-Ping~Yan$^{\ast}$,
Wenhui~Yu,
Jing~Ye$^{\ast}$,
Siming~Liu,
Yang~Su,
Huirong~Yan,
Ping~Zhang\\
Guotianci~Xu,
Jun~Lin,
Haisheng~Ji,
Zongjun~Ning,
Wei~Chen,
Li~Ji,
Jingxing~Wang\\
\small$^\ast$Corresponding author. Email: yanshuping@pmo.ac.cn; yj@ynao.ac.cn\\
\end{center}

\subsubsection*{This PDF file includes:}
Materials and Methods\\
Supplementary Text\\
Figures S1 to S16\\
Tables S1 to S4\\

\subsubsection*{Other Supplementary Materials for this manuscript:}
Movies S1 to S18\\

\newpage


\subsection*{Materials and Methods}

\subsubsection*{AIA Imaging Data and Processing}

Ultraviolet imaging observations of the solar flare were obtained from the Atmospheric Imaging Assembly (AIA) aboard the Solar Dynamics Observatory (SDO)~\cite{Lemen2012,Pesnell2012}. The AIA images the solar atmosphere in eight ultraviolet narrow-band channels plus one continuum band, with a spatial resolution of 1.5~arcseconds and a cadence of 12~seconds~\cite{Lemen2012}. Level~1.0 data were processed to Level~1.5 using standard \texttt{SolarSoft} routines, including \texttt{read\_sdo.pro} and \texttt{aia\_prep.pro}, to correct instrumental artifacts and co-align images across channels.

To quantify the distribution of plasma emission measure as a function of temperature, we performed differential emission measure (DEM) analysis using the sparse inversion technique~\cite{Cheung2015,Su2018} on six AIA channels (94, 131, 171, 193, 211, and 335~\AA). Prior to inversion, the data were rebinned by a factor of $2 \times 2$ pixels to improve the signal-to-noise ratio.

\subsubsection*{RHESSI Imaging Data and Processing}

X-ray imaging and light curve data were obtained from the Reuven Ramaty High Energy Solar Spectroscopic Imager (RHESSI)~\cite{Lin2002}. RHESSI employed nine rotating modulation collimators coupled with high-resolution germanium detectors, achieving an angular resolution of $\sim$2.3~arcseconds and an energy resolution of $\sim$1~keV FWHM at low energies. For reliable image reconstruction, we used detectors 3--6 and 8F, which provided adequate modulation.

Based on the RHESSI 25--50~keV light curve and data quality, we selected three time intervals for imaging analysis. The first interval captures a minor peak during the flare's precursor phase. The second and third intervals correspond to the main Hard X-ray (HXR) peak, split into early and late peaks due to a change in the RHESSI attenuator state (from A1 to A3). Each image was integrated over $\sim$50~seconds to ensure sufficient photon counts. We applied multiple imaging algorithms---CLEAN, PIXON, and the Expectation-Maximization method~\cite{Hogbom1974,Metcalf1996,Benvenuto2013}---all of which yielded consistent source morphologies. The figures present images reconstructed with the PIXON algorithm, chosen for its clear revelation of morphological structure and superior photometric accuracy~\cite{Aschwanden2004}.

HXR source centroids for the three intervals were derived using a weighted average of pixel brightness within each source region (see \texttt{map\_xymoments.pro}); centroid uncertainties were estimated via the visibility forward-fit algorithm~\cite{Hannah2008} and are provided in table~\ref{tab:centroids}. Distances between HXR and soft X-ray sources were calculated based on these centroids. We adopted a solar radius of 949~arcseconds based on AIA observations and defined the flare's reference position as the cusp base at helioprojective Cartesian coordinates (715, --315)~arcseconds; the cusp centroid itself is located at approximately (716, --313)~arcseconds.

For spectroscopic analysis, we used RHESSI detector 3F and analyzed spectra with the OSPEX software package~\cite{Schwartz2002}. Following ref.~\cite{Zhang2018}, we applied two models to fit the RHESSI spectra. The first combined a thermal component (isothermal bremsstrahlung based on CHIANTI~\cite{Landi2006}) with a non-thermal thick-target component~\cite{Brown1971}. The second used two thermal (adiabatic) components. Both models provided satisfactory fits (fig.~\ref{fig:RHESSIspec}).

\subsubsection*{Nobeyama Radio Observations}

Microwave data were obtained from the Nobeyama Radio Polarimeters (NoRP) and Radioheliograph (NoRH) at the Nobeyama Radio Observatory, Japan~\cite{Nakajima1994}. NoRP continuously observes the full Sun at 1--80~GHz with both cirlar polarizations, providing high-cadence light curves. NoRH is a T-shaped interferometer of 84~80-cm antennas that produces full-disk solar images at 17~GHz (dual polarization) and 34~GHz (intensity only), with $\sim$10'' and $\sim$5'' angular resolution, respectively, and 1~s temporal resolution (0.1~s for flare events).

The microwave imaging data used in this study, from 27~August~2015 at 05:37:35~UT, include 17~GHz Stokes $I$ and $V$ (fi17, fv17) and 34~GHz Stokes $I$ (fi34). All data were processed and analyzed using the NoRH Software Library, which provides routines for reading, calibrating, and analyzing interferometric data (fig.~\ref{fig:NoRH}). To enhance the signal-to-noise ratio, we applied a statistical thresholding technique: a quiet-Sun region was selected to calculate the mean ($\mu$) and standard deviation ($\sigma$) of the background emission, and pixels with values below $\mu + 3\sigma$ were masked. For coherent spectral analysis, the higher-resolution 34~GHz data were convolved to match the 17~GHz beam.

The energy distribution of accelerated nonthermal electrons is described by a power law, $dN(E)/dE \propto E^{-\delta}$, where $\delta$ is the nonthermal spectral index. We derived $\delta$ from the radio spectral index $\alpha$ using the empirical relation of ref.~\cite{Dulk1985}:

\begin{equation}
\delta = \frac{1.22 - \alpha}{0.895}, \qquad 
\alpha = \frac{\log(S_{34}/S_{17})}{\log(\nu_{34}/\nu_{17})},
\end{equation}
where $S$ is the flux density and $\nu$ is the frequency.

The degree of circular polarization at 17~GHz, $r_c = \text{fv17} / \text{fi17}$, provides information on the magnetic field direction. Following ref.~\cite{Dulk1985}, the magnetic field strength $B$ was derived from:

\begin{equation}
|r_c| = 1.26 \times 10^{0.035\delta} 10^{- 0.071 \cos\theta} \cdot \left(\frac{\nu}{\nu_B}\right)^\xi,
\end{equation}
where $\theta$ is the angle between the magnetic field and the line of sight, $\nu_B = eB/(2\pi m_e c) \approx 2.8 \times 10^6 B$ (GHz) is the electron gyrofrequency, and $\xi = -0.782 + 0.545\cos\theta$.

Rearranging gives:

\begin{equation}
\frac{\nu}{\nu_B} = \left( \frac{|r_c|}{1.26 \times 10^{0.035\delta - 0.071 \cos\theta}} \right)^{1/\xi},
\qquad
B = \frac{\nu}{2.8 \times 10^6 \cdot (\nu/\nu_B)}.
\end{equation}

For this flare, using a solar radius of 949~arcseconds and a reference position of (715, --315)~arcseconds (derived from AIA analysis), we obtained a line-of-sight angle $\theta = 55.4^\circ$ and $1/\xi \approx -2.116$.

The gyrosynchrotron formula in ref.~\cite{Dulk1985} can only detect magnetic fields within $\sim$60--600~G, a limitation arising from the interplay of observational band constraints, electron properties, and underlying physical assumptions. This range encompasses the typical magnetic field strengths of solar active regions, ensures the validity of the model's assumptions (mildly relativistic electrons, optically thin, uniform source), and is well matched to the sensitivity of 17--34~GHz radio telescopes. Fields outside this range either produce undetectably weak emission or violate the model's core premises, precluding reliable determination.

\subsubsection*{Wavelet Analysis}

Wavelet analysis was performed on light curves from NoRP (2, 3.75, and 9.4~GHz), NoRH (17~GHz), AIA (summed 171, 211, 304, 335, and 1600~\AA), RHESSI (25--50~keV), and on simulated flare parameters. Gaps in the NoRP light curves were filled using shape-preserving piecewise cubic interpolation, with random noise added at the level of the local standard deviation to preserve signal variability. The 2, 3.75, and 9.4~GHz NoRP light curves were co-added to improve the signal-to-noise ratio. Time resolutions were 10~s (NoRP and NoRH), 12~s (AIA), 4~s (RHESSI), and 9.67~s (simulation). We used the wavelet software package of ref.~\cite{Torrence1998} with the Morlet mother wavelet to optimize frequency resolution~\cite{Moortel2004}.

\subsubsection*{Parameter-optimized Hilbert-Huang Transform Analysis}

Time-frequency analysis of the flare light curves was performed using the parameter-optimized Hilbert-Huang transform (PHHT)~\cite{Yan2024}. The traditional Hilbert-Huang transform (HHT) is an adaptive technique for high-resolution analysis of nonlinear, non-stationary processes~\cite{Huang1998,Wu2009,Colominas2014}. PHHT generalizes HHT and overcomes two key limitations: difficulty extracting small-amplitude signals embedded in large-amplitude explosive events, and reliance on empirical validation. PHHT proceeds in two steps: (1) empirical mode decomposition (EMD) to obtain intrinsic mode functions (IMFs), and (2) Hilbert spectral analysis of the IMFs to generate time-frequency spectra.

Our implementation builds on the complete ensemble EMD with adaptive noise (CEEMDAN) algorithm, using modified codes from \url{http://perso.ens-lyon.fr/patrick.flandrin/emd.html}. The key modification is to optimize the noise amplitude at each decomposition stage according to the orthogonality index, thereby maximizing the orthogonality among the IMFs~\cite{Yan2024}. The optimization minimizes both the overall orthogonality index ($IO$) and the pairwise indices between neighboring IMFs ($IO_{fg}$)~\cite{Huang1998}:

\begin{equation}
    IO = \sum_{t=0}^{T} \left( \sum_{j=1}^{n+1} \sum_{k=1}^{n+1} \frac{C_{j}(t)C_{k}(t)}{X^{2}(t)} \right),
\end{equation}

\begin{equation}
    IO_{fg} = \sum_{t} \frac{C_{f}(t)C_{g}(t)}{C_{f}^{2}(t) + C_{g}^{2}(t)},
\end{equation}

\noindent where $t$ is time, $T$ is the total duration, $C_j(t)$ is the $j$-th EMD component (IMF or residual), $X(t) = \sum_{j=1}^{n+1} C_j(t)$ is the original signal, and $n$ is the number of IMFs.

We used the default sifting stopping criterion~\cite{Rilling2003} with 2000 realizations, a maximum of 5000 sifting iterations, and constant signal-to-noise ratio across all stages. After decomposition, the normalized Hilbert transform~\cite{Huang2009} was applied to obtain instantaneous frequencies and amplitudes for each IMF.

Although HHT-derived instantaneous frequencies generally reflect underlying physics, the method does not guarantee physically consistent instantaneous frequencies in all cases~\cite{Huang1998}. To ensure robust analysis, we applied quality thresholds: $IO < 0.05$ for AIA/RHESSI data and $IO < 0.01$ for NoRP/NoRH and simulation data. Spectra exhibiting fragmentary structures, abnormal time-frequency features, or negative frequencies in the main frequency band were excluded. The remaining high-quality spectra were averaged to derive mean frequency curves, with standard deviations quantifying frequency uncertainty.

\subsubsection*{Cross-Correlation between QPOs}

During the flare's pulse phase, the time-frequency profiles of QPOs in the microwave, ultraviolet, and HXR bands exhibit similar characteristics (Fig.~\ref{fig:QPPdatasim}). To quantify the time lags between them, we implemented a Monte Carlo cross-correlation approach. Because the original time resolutions differed across instruments, all curves were interpolated to 1~s using spline interpolation. For each cross-correlation analysis (observed and simulated), we performed $10^6$ Monte Carlo realizations in which each curve was perturbed with random noise at the level of its measured standard deviation. The resulting cross-correlation functions were fitted with a composite model (Gaussian plus quadratic polynomial) to determine the lags, and the lag distributions were then fitted with Gaussians to obtain the mean lag and its standard deviation.

\subsubsection*{Cross-Correlation between Light Curves}

Cross-correlation analysis was performed on the HXR, microwave, and ultraviolet light curves. All curves were interpolated to a uniform 1~s resolution using spline interpolation. To assess the robustness of the temporal lag between the HXR and microwave light curves, we systematically varied the start and end times of the analysis interval in 1~s steps and computed the cross-correlation coefficient as a function of lag for each interval (fig.~\ref{fig:LCcorrelate}). The resulting cross-correlation functions were fitted with a model consisting of a Gaussian component plus a linear baseline to determine the lags. The final lag and its uncertainty are, respectively, the mean and standard deviation of the lags derived from all intervals (fig.~\ref{fig:lags2eng}). The lag between the ultraviolet and microwave light curves, along with its standard deviation, was calculated using a focused interval containing the main burst to minimize contamination from an earlier ultraviolet pre-burst.

\subsubsection*{Flare Magnetohydrodynamic Simulation}

To enable direct comparison with observational data, we performed a 2.5-dimensional magnetohydrodynamic simulation based on the standard CSHKP flare model\cite{Kopp1976}. Our model incorporates the effects of resistivity, gravity, anisotropic thermal conduction, radiative cooling, and background coronal heating. The governing equations are as follows:
\begin{eqnarray}
&&\frac{\partial \rho}{\partial t} + \nabla\cdot(\rho\textbf{v})=0,\label{mhd1}\\
&&\frac{\partial e}{\partial t}+\nabla\cdot\left[(e+P^* )\textbf{v}-(\textbf{v}\cdot \textbf{B})\textbf{B}\right]\nonumber\\
&&=\rho\textbf{g}\cdot\textbf{v}+\nabla\cdot\left[ \eta\textbf{B}\times (\nabla\times\textbf{B})+\kappa_{||}(\nabla T\cdot \hat{\textbf{B}})\hat{\textbf{B}}\right]-n_in_eQ(T)+H\label{mhd2},\\
&&\frac{\partial(\rho \textbf{v})}{\partial t}+\nabla\cdot\left[\rho\textbf{v}\textbf{v}-\textbf{BB}+P^*\textbf{I}\right]=\rho\textbf{g},\label{mhd3}\\
&&\frac{\partial\textbf{B}}{\partial t}=\nabla\times(\textbf{v}\times \textbf{B}-\eta\nabla\times\textbf{B}),\label{mhd4}\\
&&P^* =P+\frac{\textbf{B}\cdot \textbf{B}}{2}, \label{mhd5}\\
&&P=\rho T,\label{mhd6}
\end{eqnarray}
where $\rho$, $\mathbf{v}$, $P$, $T$, $\mathbf{B}$, and $\hat{\mathbf{B}}$ denote the mass density, flow velocity, thermal pressure, temperature, magnetic field, and unit magnetic vector, respectively. The total energy density is given by $e = P/(\gamma-1) + \rho\mathbf{v}^2/2 + \mathbf{B}\cdot\mathbf{B}/2$, where $\gamma = 5/3$ is the ratio of specific heats for an ideal gas. The heat flux is computed as $F_C = \kappa_{\parallel}(\nabla T\cdot\hat{\mathbf{B}})\hat{\mathbf{B}}$, with the Spitzer conductivity\cite{Spitzer1962} given by $\kappa_{\parallel} = 1.84 \times 10^{-10} T^{5/2}/\ln\Lambda$ in SI units, where the Coulomb logarithm is fixed at $\ln\Lambda = 30$. The radiative cooling rate $Q(T)$ follows the piecewise function for fully ionized hot plasma described by ref.\cite{Klimchuk2008}. The heating function $H = n_in_eQ(T_c)$ maintains thermal balance at the coronal temperature $T_c = 2\times 10^6$ K. These equations are solved in dimensionless form, with normalization units provided in table~\ref{mhd}. Additionally, the divergence-free condition ($\nabla\cdot\mathbf{B} = 0$) is enforced throughout the simulation to ensure numerical consistency.

The simulation domain spans $(x,y) \in [-1,1] \times [0,2]$ in non-dimensional Cartesian coordinates. We initialize a Harris current sheet configuration with uniform resistivity $\eta = 10^{-6}$. The initial magnetic field is prescribed as:
\begin{equation}
B_x = 0, \quad B_y =
\begin{cases}
\sin\left(\frac{\pi x}{2\lambda}\right), & |x| \leq \lambda \\
1, & x > \lambda \\
-1, & x < -\lambda
\end{cases}, \quad
B_z = \sqrt{1 - B_y^2},
\end{equation}
with current sheet half-width $\lambda = 0.1$. To initiate fast magnetic reconnection, we introduce a localized perturbation to the magnetic potential:
\begin{equation}
A_{\epsilon} = \epsilon \cdot \exp\left[-\left(\frac{x}{l_x}\right)^2 - \left(\frac{y - y_c}{l_y}\right)^2\right],
\end{equation}
with parameters $\epsilon = 0.01$, $y_c = 0.5$, $l_x = 0.05$, and $l_y = 0.05$.

The model incorporates a gravitationally stratified atmosphere with a smooth chromosphere-to-corona temperature transition:
\begin{equation}
T(y) = \frac{T_{\mathrm{cor}} - T_{\mathrm{chr}}}{2} \tanh\left(\frac{y - h}{w}\right) + \frac{T_{\mathrm{cor}} + T_{\mathrm{chr}}}{2},
\end{equation}
where $T_{\mathrm{cor}} = 7.71 \times 10^{-3}$ is the coronal temperature and $T_{\mathrm{chr}} = 1.66 \times 10^{-5}$ represent the chromospheric temperature. The height and width of the transition region are $h = 0.03$ and $w = 0.0075$.

The density profile is determined by hydrostatic equilibrium:
\begin{equation}
	\nabla P = -\rho \bf{g},
\end{equation}
where the normalized gravitational acceleration is given by
\begin{equation}
\mathbf{g} = \frac{-g_0 \hat{\mathbf{y}}}{(1 + yL_0/R_\odot)^2} \cdot \frac{\rho_0 L_0}{P_0},
\end{equation}
with solar radius $R_\odot = 6.691 \times 10^8$ m and gravitational constant $g_0=274$ ms$^{-2}$. The $\hat{\textbf{y}}$ is the unit vector along the $y$-axis. 

Simulations were performed using the open-source ATHENA framework\cite{Stone2008} on a uniform $3840 \times 3840$ grid, corresponding to a spatial resolution of 63 km. The full magnetohydrodynamic equations are solved using an unsplit Godunov method with an HLLD Riemann solver\cite{Miyoshi2005} to accurately capture plasma dynamics and shock structures. Anisotropic thermal conduction is treated using a novel semi-implicit method developed by ref. \cite{Ye2020b}. Boundary conditions consist of line-tied photospheric boundary at $y=0$ and open boundaries elsewhere. While our previous work\cite{Ye2020} focused on extreme-ultraviolet QPOs from plasmoid collisions, this study specifically investigates the mechanisms underlying HXR, microwave, and ultraviolet QPOs.


\subsection*{Supplementary Text}

\subsubsection*{Detailed flare evolution in the main phase}

The flare began with a whirling plasma ejection (fig.~\ref{fig:CSevolution}A), which stretched a heart-shaped magnetic structure (fig.~\ref{fig:CSevolution}, B and C) and formed a current sheet between them (fig.~\ref{fig:CSevolution}, E and F). A non-thermal HXR source was precisely located between the stretching heart-shaped structure and the ejected material (fig.~\ref{fig:CSevolution}, C and D), indicating efficient particle acceleration. Meanwhile, an anchored magnetic loop underwent progressive twisting and stretching, culminating in a clear rupture and reconnection (fig.~\ref{fig:CSevolution}, D and E; movie~1). This topological reorganization was accompanied by the emergence of a second HXR source located below the first (fig.~\ref{fig:CSevolution}E). The reconnection produced a classic cusp-shaped loop topped by a plasma sheet that spatially coincided with the first HXR source (fig.~\ref{fig:CSevolution}F). A plasma sheet is essentially a plasma-filled current sheet. This confirms that electrons were accelerated directly within the current sheet before being injected into the lower atmosphere to generate the second HXR source. Later, the cusp dissipated, and the heart-shaped structure re-emerged with a more symmetric morphology (fig.~\ref{fig:CSevolution}, G and H), indicating a relaxed post-reconnection state. AIA 94~\AA~imaging further confirms that this heart-shaped structure was embedded within a highly twisted flux rope that passed through several low-lying parallel loops (fig.~\ref{fig:CSevolution}I).

\subsubsection*{QPO–B Power-Law Scaling across astrophysical systems}

To test whether twisted-loop magnetic reconnection powers outbursts from black hole X-ray binaries, active galactic nuclei, magnetars, and gamma-ray bursts, we compiled their QPO frequencies and magnetic field strengths from the literature (table~\ref{QBsource0}), associating each QPO with its corresponding field strength (table~\ref{QBsource1}).

QPOs are frequently observed in solar flares~\cite{McLaughlin2018,Zimovets2021,Corchado2024}. To reliably relate magnetic field strength to QPO frequency specifically for reconnection-region oscillations, we restrict our analysis to unambiguous QPO signatures from current sheets. We therefore focus on two flares: SOL2015-08-27T05:45 from this study and SOL2017-07-13T21:56 from ref.~\cite{Kou2022}.

For flare SOL2015-08-27T05:45, the observed QPO frequency ranges from $\sim$5.14 to $\sim$7.37~mHz (Fig.~\ref{fig:QPPfre}). The magnetic field strength in the reconnection region falls below the detection limit of $\sim$60~G (Fig.~\ref{fig:NoRH}C). Since magnetohydrodynamic simulations yield QPO frequencies consistent with the observations (Fig.~\ref{fig:QPPdatasim}, C and F), we adopt the simulated field strengths of $\sim$10--25~G as the counterparts to the observed QPO frequencies (Fig.~\ref{fig:Simulpara}I). A key feature of the hard X-ray QPO is that it appears predominantly during the precursor and rising phases of the hard X-ray outburst and diminishes during the decay phase (Fig.~\ref{fig:QPPdatasim}A).

For flare SOL2017-07-13T21:56, wavelet transforms of the brightness temperatures at 3.4 GHz and 8.4 GHz detect two QPOs: one with a period of $\sim$10--20~s ($\sim$50--100~mHz) and another with a period of $\sim$30--60~s ($\sim$16.7--33.3~mHz). In the current sheet region, the 30--60~s QPO persists for over two minutes and achieves high significance in the power spectrum, while the 10--20~s QPO lasts only approximately one minute with low significance. Consequently, we identify the 30--60~s QPO as the signature of magnetic-reconnection-generated oscillations. For comparison with observationally derived results, the MHD numerical simulations presented in ref.~\cite{Kou2022} employ a magnetic field strength of 40 G. Accordingly, we estimate the magnetic field strength associated with this QPO to be approximately 40 G. The lack of hard X-ray observations makes it difficult to precisely determine the phase during which this QPO is generated.

QPOs are common in the X-ray emission of black hole X-ray binaries and are broadly divided into low-frequency ($\lesssim$30~Hz) and high-frequency ($\gtrsim$60~Hz) types. Low-frequency QPOs are further classified into types A, B, and C. Type-C QPOs are the most common and appear across all accretion states. Their frequency rises from a few mHz in the hard state to $\sim$10~Hz in intermediate states and occasionally reaches $\sim$30~Hz in soft or ultra-luminous states. Type-B QPOs (5--6~Hz) occur in the soft intermediate state and are temporally linked to jet ejections. Type-A QPOs (6--8~Hz) appear in the soft state just after the hard-to-soft transition. High-frequency QPOs are only seen during high-flux episodes. However, the physical origin of all these QPOs remains controversial~\cite{Ingram2019}.

Accretion disks can generate and amplify magnetic fields, forming twisted magnetic loops that undergo reconnection, thereby driving the observed spectral evolution~\cite{Romanova1998, Donati2005, Matteo1999, Jacquemin2024}. If QPOs originate from magnetic reconnection and their frequency scales with the magnetic field strength, the observed diversity follows naturally. The frequency range of Type-C QPOs would then trace the systematic increase of the magnetic field as the accretion flow evolves from hard to soft states. The stable 5--6~Hz of Type-B QPOs may mark a critical field strength for jet ejection, consistent with their association with radio flares. Type-A QPOs could reflect a new magnetic equilibrium in the disk after the state transition. High-frequency QPOs would require extreme fields in the innermost region, possibly involving higher harmonics of the tearing mode, which can produce fundamental and harmonic frequencies in reconnecting current sheets~\cite{Somov1993}.

On the other hand, during the early rising phase of MAXI~J1820+070 in the bright hard state, it exhibits intermittent, low-frequency ($<$1~Hz) QPOs that alternate between active and quiescent episodes~\cite{Zhang2023}. This intermittency is naturally explained by magnetic reconnection: the weak disk magnetic field in the early outburst yields a low QPO frequency and few magnetic loops. The resulting long intervals between successive reconnection events produce intermittent QPOs. Conversely, when accretion strengthens the disk magnetic field, the intervals become short or overlapping, yielding continuous QPOs.

We therefore select data from black hole X-ray binaries to investigate the QPO--B relationship. Table~\ref{QBsource0} lists all black hole X-ray binaries with both QPO and magnetic field estimates, along with several other typical black hole X-ray binary QPO sources. 
 
For the black hole transient MAXI J1820+070, the QPO frequency decreased from 9.7~Hz to 1.1~Hz between MJD~58386.2 and MJD~58392.1~\cite{Bellavita2025}. Meanwhile, the magnetic field strength in the accretion disk evolved from $\sim$$2.4\times10^5$--$5\times10^7$~G to $\sim$$3\times10^4$--$7\times10^7$~G between MJD~58389 and MJD~58397~\cite{You2023}. We therefore adopt the QPO frequency of 2.96~Hz at MJD~58389.156 and the magnetic field strength of $\sim$$2.4\times10^5$--$5\times10^7$~G at MJD~58389 as the corresponding data point. 

For GX~339--4, QPOs at frequencies of 0.2--8~Hz have been detected~\cite{Belloni2005}, and the magnetic field strength of the accretion disk has been estimated to be $\sim$$10^{8.06}$~G~\cite{Daly2019}. If magnetic reconnection occurs within a magnetic loop in the accretion disk, the reconnection site is expected to be located above the disk plane, where the magnetic field strength is lower than that within the disk. Furthermore, during the early rising phase of an outburst, the disk magnetic field strength is also expected to be lower than this estimated value. Based on the radial profile of the accretion disk magnetic field presented in~\cite{You2023}, which shows a decrease of approximately two orders of magnitude from the inner to the outer disk, we infer that the field strength associated with the 0.2--8~Hz QPOs lies in the range of $10^{6.06}$--$10^{8.06}$~G. 

The infrared-emitting region of a jet is likely located farther from the magnetic reconnection site, where the magnetic field strength is expected to be lower than that in the reconnection region. We therefore do not adopt the value of $1.5 \times 10^4$~G derived for GX~339--4 from mid-infrared jet observations~\cite{Gandhi2011}.

Abundant QPO data exist for XTE~J1550-564, GRS~1915+105, XTE~J1859+226, and GRO~J1655-40~\cite{Belloni2012,Cui1999,Li2013,Morgan1997,Yan2013,Casella2004,Motta2012}. However, the inferred magnetic field strength of $5 \times 10^4$~G for XTE~J1550-564, which is based on near-infrared jet emission~\cite{Chaty2011}, is not adopted for the same reason. Similarly, magnetic field strengths in accretion disk winds are expected to be lower than those in the reconnection region; consequently, we do not adopt the value of $10^{3}$--$10^{5}$~G derived for GRS~1915+105 from disk wind observations~\cite{Miller2016}. For XTE~J1859+226 and GRO~J1655-40, no magnetic field strength estimates are available.

QPOs at millihertz or even lower frequencies have been detected in three active galactic nuclei and two tidal disruption events~\cite{Masterson2025,Gierlinski2008,Lin2013,Pasham2019,Reis2012}. Among these sources, only 1ES 1927+654 has a corresponding magnetic field estimate~\cite{Laha2025,Meyer2025}. In 1ES~1927+654, a QPO at $\sim$0.93 mHz was detected in July 2022, which increased to $\sim$2.34 mHz by March 2024, accompanied first by radio (5 GHz) flaring and then by a resolved two-sided jet. The jet is resolved at 0.1--0.3 pc, with a magnetic field strength of $\sim$0.2 G measured at 0.1 pc, while the QPO originates from the hot corona. We find that magnetic reconnection generates hard X-rays and microwaves (2--17 GHz) with QPO signatures and simultaneously produces or heats the corona. This suggests that the mHz QPOs in 1ES~1927+654 may arise from a twisted magnetic loop, and the field strength of the loop is comparable to the $\sim$0.2 G measured at the jet base. In this picture, the rising QPO frequency traces an increasing field strength. Such an increase in the inner accretion flow magnetic field with ongoing accretion has been observed in MAXI J1820+070~\cite{You2023}. The persistence of the QPO over more than two years further implies that magnetic loops are successively formed and undergo reconnection during this period.

Magnetic loop reconnection is widely considered the driver of outbursts from magnetars and gamma-ray bursts~\cite{Thompson2002, Parfrey2013, Kaspi2017, Carrasco2019, Lyubarsky2020, Mahlmann2023, Most2022, Most2023}. QPOs have been detected in three magnetars---SGR~1806-20, SGR~1900+14, and SGR~J1935+2154---all with estimated magnetic field strengths. In SGR~1806-20, a 626.5~Hz QPO has been detected in the hard X-ray band (25--100~keV)~\cite{Watts2006}. This QPO appears during the precursor and rising phases of the outburst, analogous to the solar flare QPOs in this study, suggesting a common magnetic reconnection origin. The dipolar magnetic field strength of SGR~1806-20 is estimated at $8 \times 10^{14}$~G~\cite{Kouveliotou1998}. The 626.5~Hz QPO likely originates in the magnetosphere, where the field strength is expected to be lower than the dipolar value. Assuming again a two-order-of-magnitude decrease as a lower bound, we estimate the associated magnetic field strength to be in the range of $8 \times 10^{12}$--$8 \times 10^{14}$~G.

High-frequency QPOs at 720, 976, 1840, and 2384~Hz, along with several lower-frequency QPOs, has also been detected~\cite{Strohmayer2006, Israel2005}. However, the 720, 976, 1840, and 2384~Hz QPOs appear only transiently (see fig.~10 in ref.~\cite{Strohmayer2006}). The 2384~Hz QPO may arise from reconnection of transient magnetic loops emerging from the magnetar's surface following crustal fractures. We therefore suggest that its associated magnetic field strength lies in the range of $8 \times 10^{13}$--$8 \times 10^{15}$~G.

For SGR~1900+14 and SGR~J1935+2154, only lower-frequency QPOs have been detected~\cite{Strohmayer2005, Li2022, Roberts2023}, potentially arising from reconnection of relatively weak magnetic field lines. Although dipole magnetic field estimates exist for both magnetars~\cite{Kouveliotou1998, Israel2016}, they do not allow us to constrain the field strengths associated with the observed QPOs. 

For gamma-ray burst GRB~910711, QPOs at 2649~Hz and 1113~Hz have been detected in the $>$100~keV band~\cite{Chirenti2023}. Numerical relativity simulations of twisted-loop magnetic reconnection have reproduced these QPOs, yielding a reconnection-region field strength of $\sim$$10^{15}$--$10^{17}$~G~\cite{Most2023}. Notably, the 2649~Hz QPO appears during the precursor and rising phases of the outburst, analogous to the QPO behavior observed in solar flare SOL2015-08-27T05:45 and magnetar SGR~1806-20. We therefore suggest that this QPO also originates from a reconnection current sheet, with an associated field strength of $10^{15}$--$10^{17}$~G.

For gamma-ray bursts GRB~200415A, GRB~230307A, GRB~120323A, GRB~181222B, and GRB~190606A, high-frequency QPOs have also been detected in the hard X-ray and gamma-ray bands~\cite{Castro-Tirado2021, Chen2025, Yang2025}. With the exception of GRB~230307A, whose QPO appears during the rising phase, all other QPOs occur during the precursor and rising phases of their respective bursts, suggesting that these QPOs may originate from magnetic reconnection. However, no magnetic field strength estimates are available for these bursts.



\begin{figure} 
	\centering
	\includegraphics[width=1.0\textwidth]{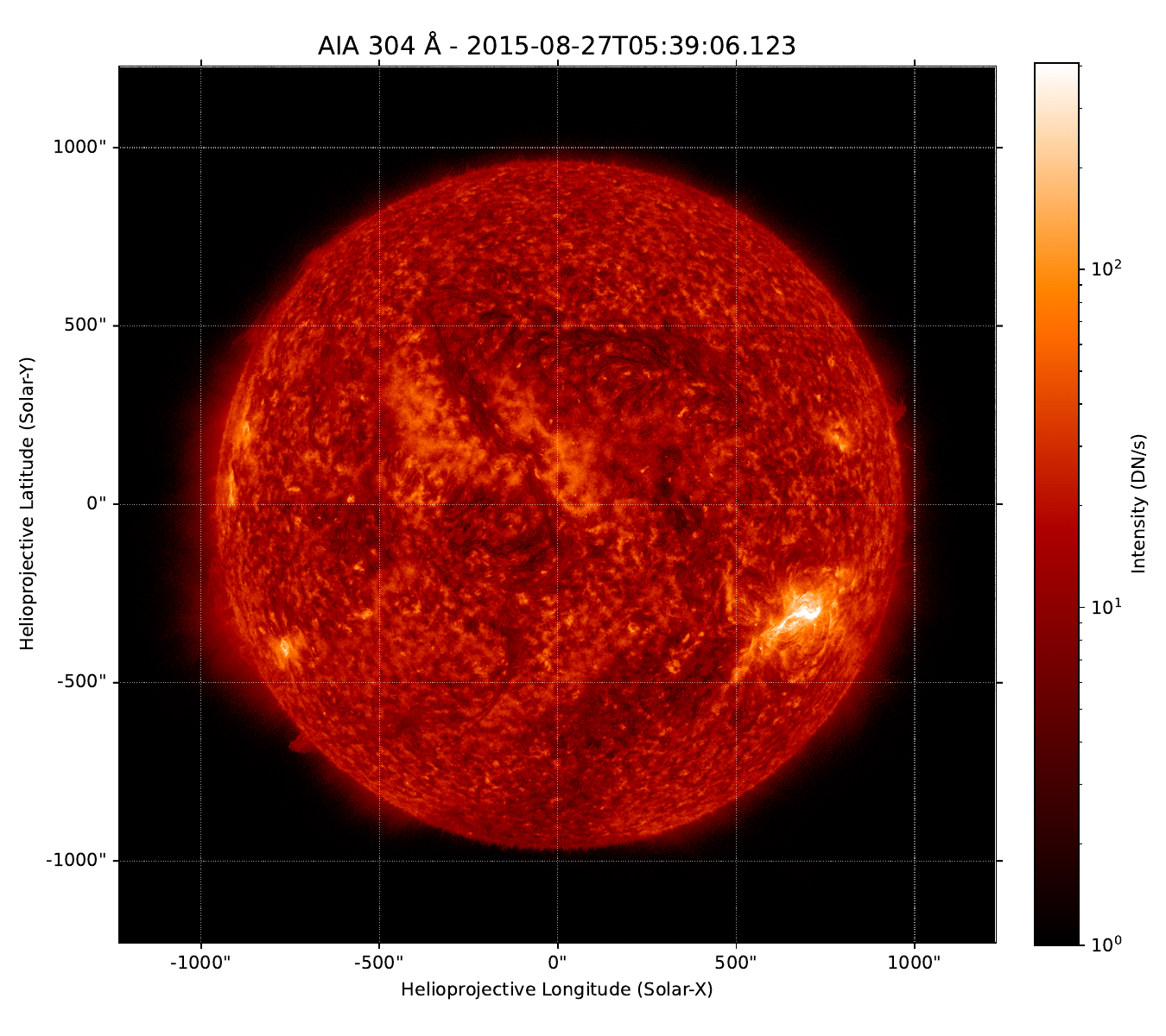} 

    \caption{\textbf{AIA~304~\AA\ full-disk image obtained during the solar flare.} 
	The solar flare is located at the bright region with Helioprojective-Cartesian coordinates of approximately (715, --315) arcseconds.}
	\label{fig:FlarePos} 
\end{figure}

\begin{figure}[htbp]
    \centering
    \includegraphics[width=1.0\textwidth]{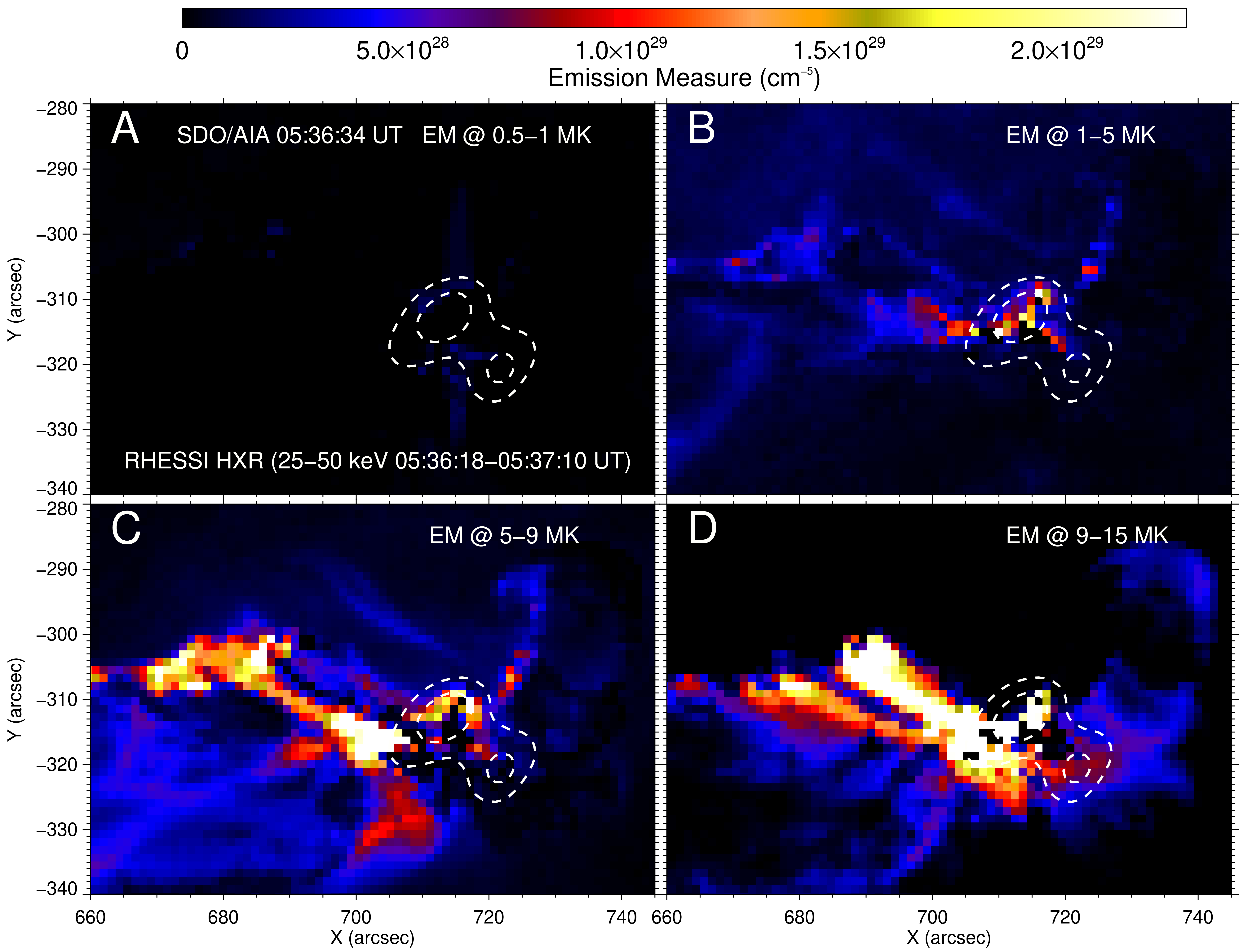}
    \caption{\textbf{Distribution of the total emission measure across four temperature intervals.}
    The four panels correspond to 0.5--1~MK (\textbf{A}), 1--5~MK (\textbf{B}), 5--9~MK (\textbf{C}), and 9--15~MK (\textbf{D}). Hard X‑ray (HXR) emission is shown as white contours (30\% and 60\% of maximum). A current sheet is clearly revealed in (B).}
    \label{fig:AIA_EM}
\end{figure}

\begin{figure}[htbp]
    \centering
    \includegraphics[width=1.0\textwidth]{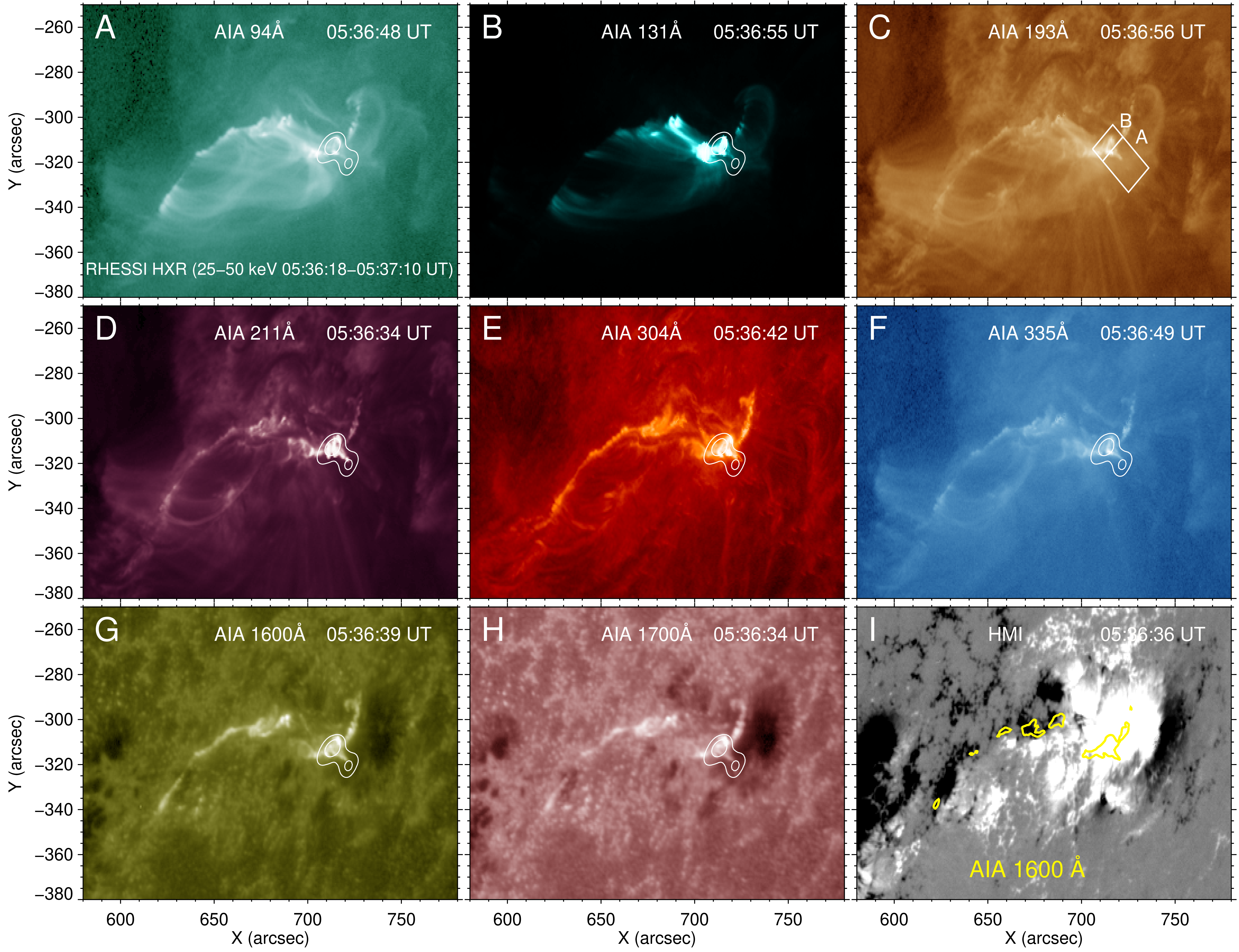}
    \caption{\textbf{The flare cusp observed in AIA wavelengths.} 
	(\textbf{A} to \textbf{H}) The cusp structure is visible simultaneously across all AIA wavelengths. The AIA~171~\AA\ image is omitted due to space constraints. Hard X-ray (HXR) emission is shown as white contours (30\% and 60\% of maximum). The two white rectangles in (C) mark the regions used to extract AIA light curves. Region A covers the cusp and the current sheet, while Region B contains the heart-shaped magnetic structure. (\textbf{I}) Photospheric magnetogram from SDO/HMI, overlaid with contours of AIA~1600~\AA\ emission.}
\label{fig:CScusp}
\end{figure}

\begin{figure}[htbp]
    \centering
    \includegraphics[width=1.0\textwidth]{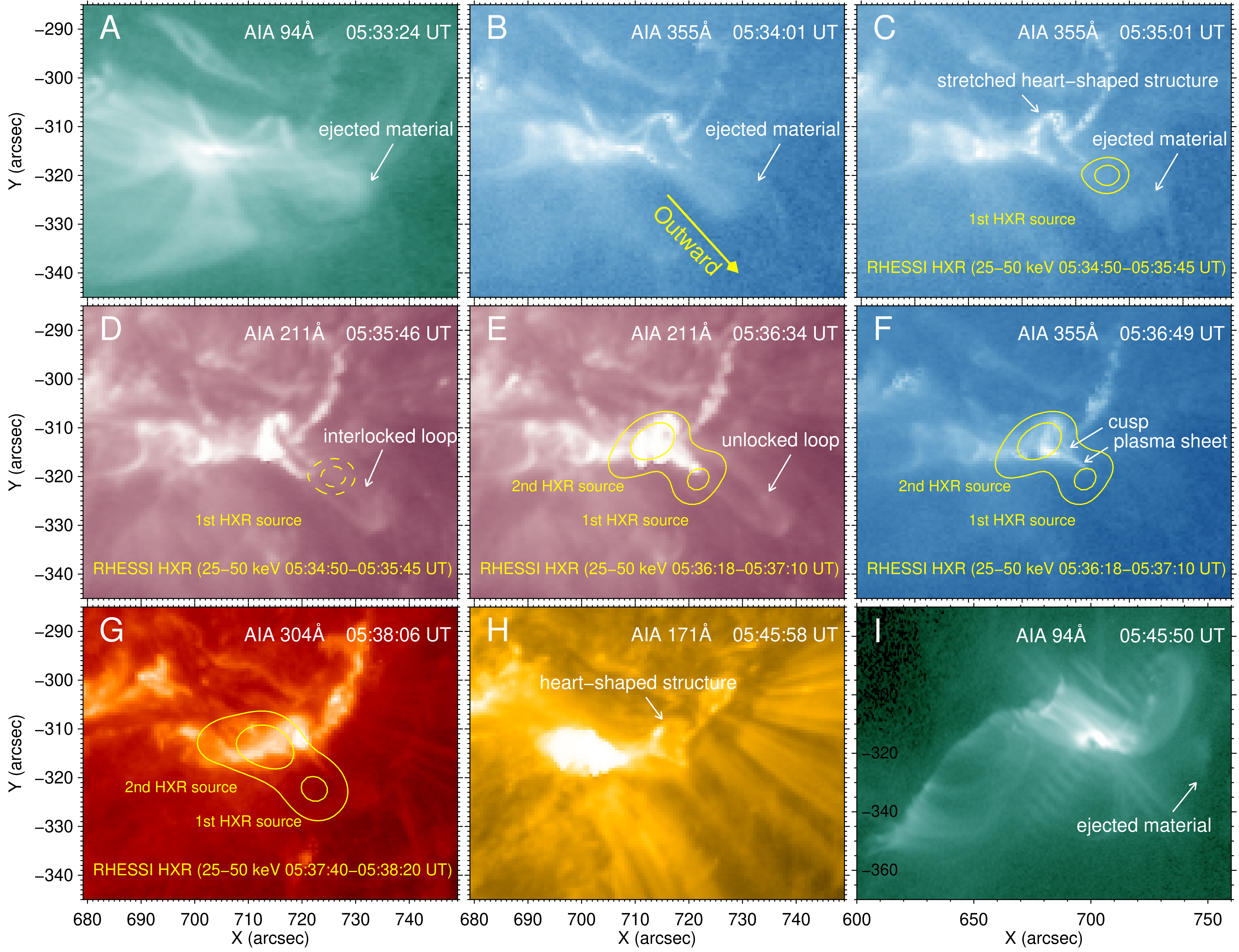}
    \caption{\textbf{Magnetic reconnection and HXR emission from current sheet.} 
	(\textbf{A}) AIA~94~\AA\ image showing a whirling ejected plasma structure. 
	(\textbf{B} and \textbf{C}) AIA~355~\AA\ images overlaid with RHESSI 25--50~keV HXR contours. A heart-shaped magnetic feature is stretched by field lines inside the ejecta, accompanied by the appearance of a HXR source between them. The yellow arrow indicates the direction outward from the sun's surface.
	(\textbf{D}) AIA~211~\AA\ image showing a magnetic loop interlocked with the heart-shaped structure. 
	(\textbf{E}) AIA~211~\AA\ image after reconnection, showing the disconnection of the loop. A second HXR source forms below the first, as indicated by RHESSI contours. 
	(\textbf{F}) AIA~355~\AA\ image displaying a cusp-shaped post-reconnection loop, with a plasma sheet situated above it. The first HXR source is co-located with the plasma sheet.
	(\textbf{G}) AIA~304~\AA\ image showing a less distinct cusp and partial reappearance of the heart-shaped structure. 
	(\textbf{H}) AIA~171~\AA\ image indicating the structure recovered to a heart-shaped one. 
	(\textbf{I}) AIA~94~\AA\ image illustrating the heart-shaped structure resided within a highly twisted flux rope that had passed through some low-lying parallel loops.}
    \label{fig:CSevolution}
\end{figure}

\begin{figure}[htbp]
    \centering
    \includegraphics[width=1.0\textwidth]{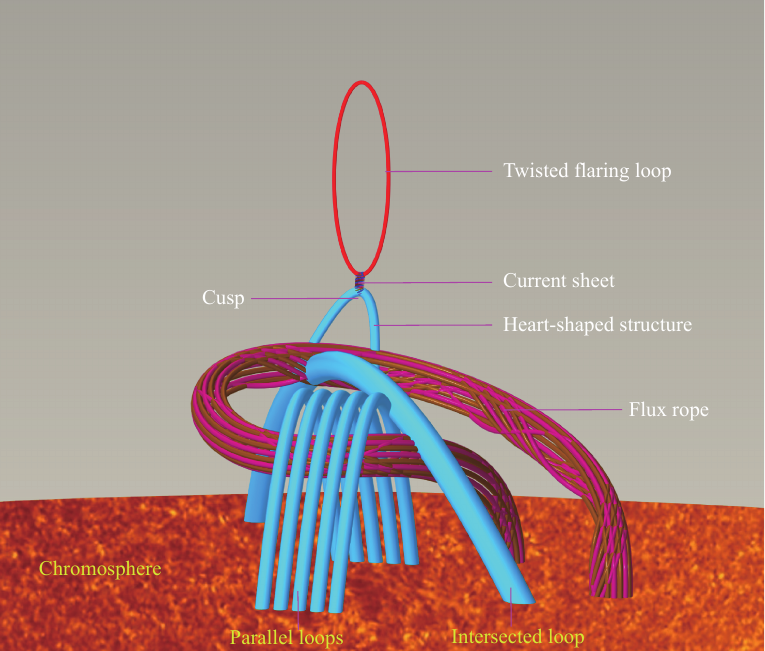}
    \caption{\textbf{Schematic of the new pattern of magnetic reconnection during the solar flare.}
	In this event, a magnetic flux rope passes through some low-lying parallel loops and intersects a larger loop, forming a heart-shaped structure. A smaller loop interlocked with this heart-shaped structure becomes twisted. This twisting motion drives the magnetic reconnection responsible for the flare, and displays classic reconnection features: a cusp and a current sheet. Magnetic loops are typically arch-like structures with a weakly non-potential magnetic field, low magnetic helicity, and relatively little stored free energy. In contrast, magnetic flux ropes are helically twisted structures characterized by a strongly non-potential magnetic field, high magnetic helicity, and a significant reservoir of free energy. 
	}
\label{fig:Flarepattern}
\end{figure}

\begin{figure}[htbp]
    \centering
    \includegraphics[width=1.0\textwidth]{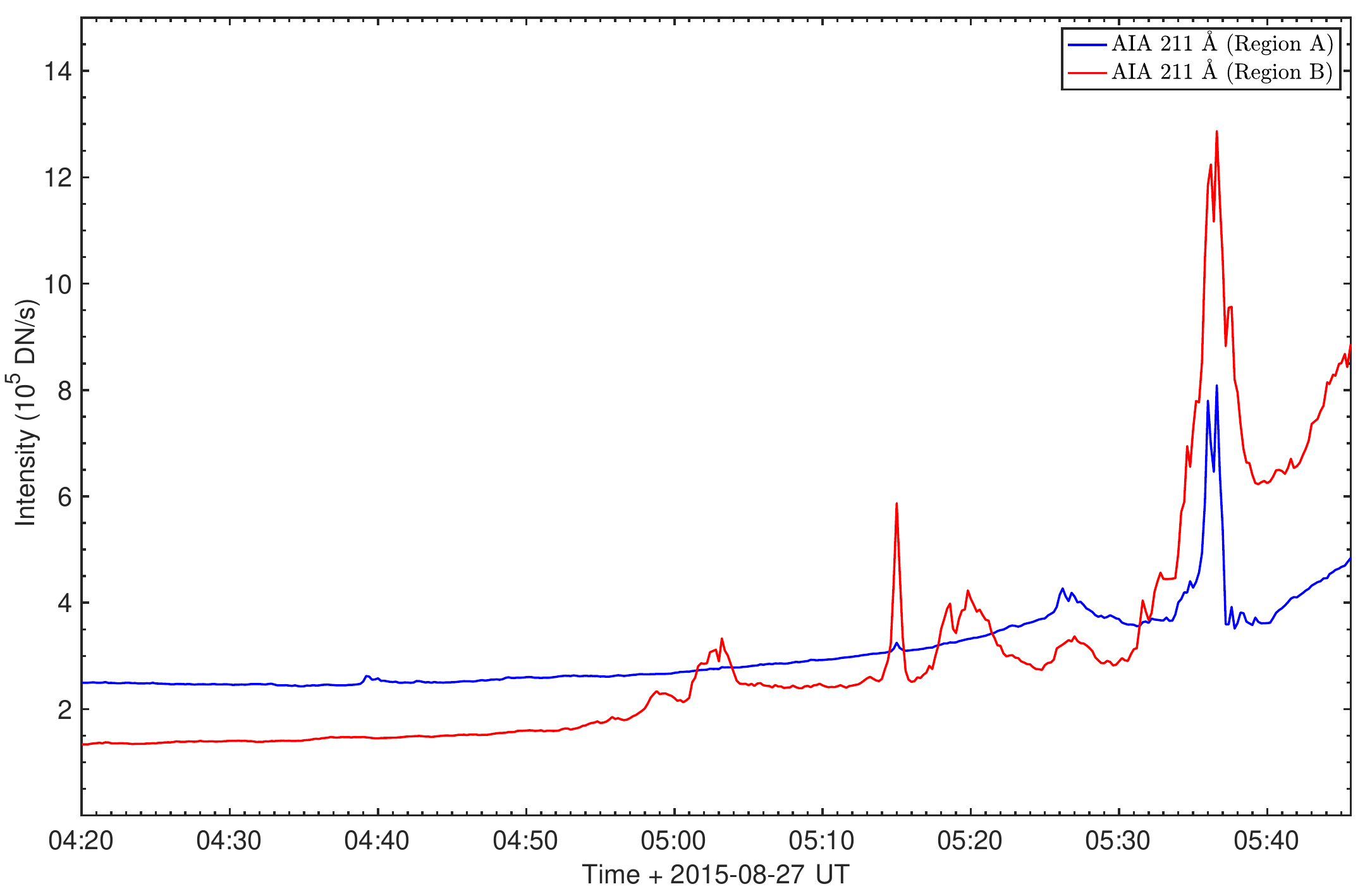}
    \caption{\textbf{AIA~211~\AA\ light curves.} The light curves are extracted from two regions (A and B) in the AIA~211~\AA\ image; the locations of these regions are shown in fig.~\ref{fig:CScusp}C.}
\label{fig:LC211}
\end{figure}

\begin{figure}[htbp]
    \centering
    \includegraphics[width=1.0\textwidth]{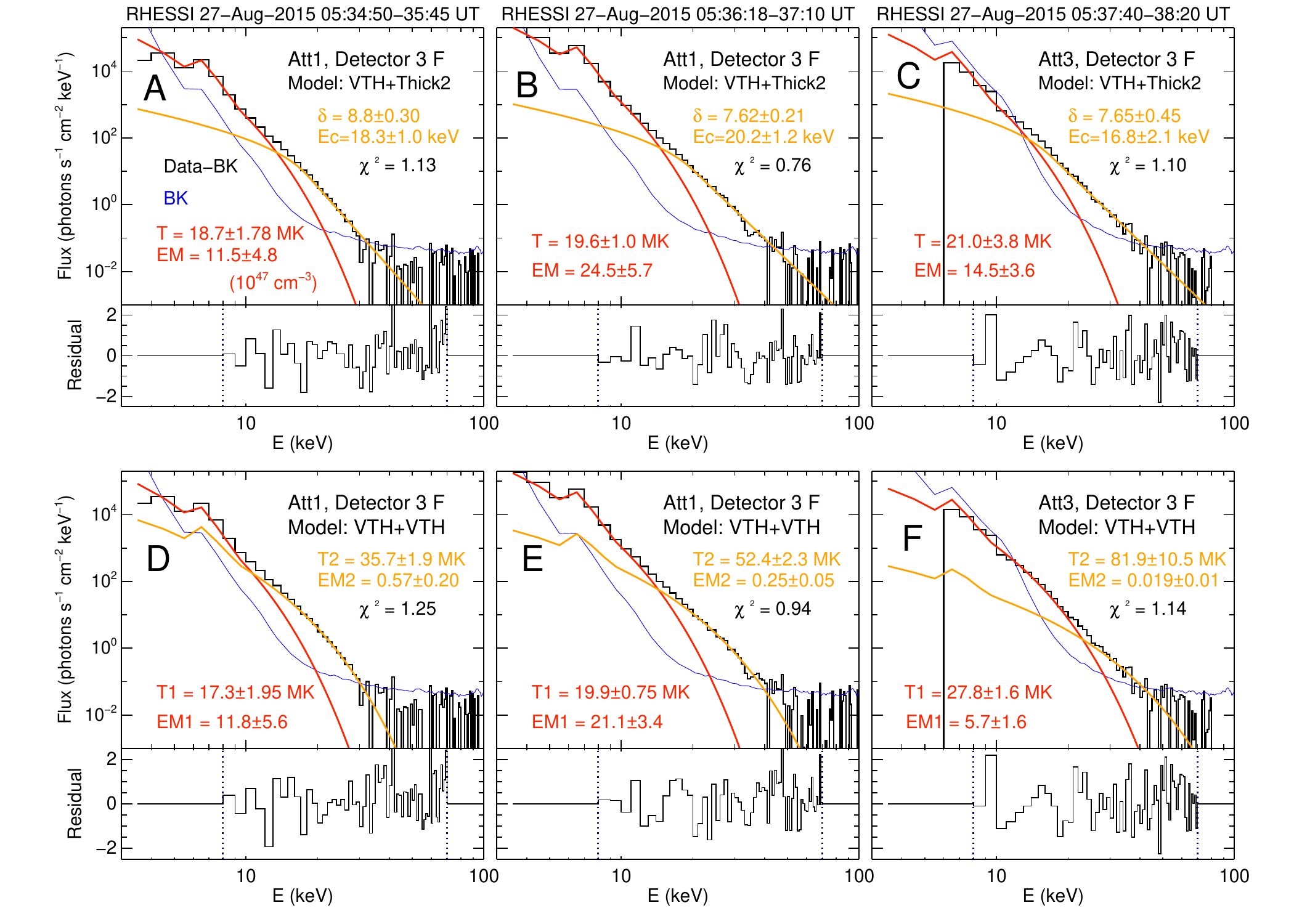}
    \caption{\textbf{Spectroscopic analysis of X-ray emission of the flare.} The X-ray spectra were fitted with two distinct models: the first model comprises a thermal (VTH) and a non-thermal (Thick2) component; the second model consists of two thermal components.
(\textbf{A} to \textbf{C}) Spectral fitting results using the first model.
(\textbf{D} to \textbf{F}) Spectral fitting results using the second model.}
\label{fig:RHESSIspec}
\end{figure}

\begin{figure}[htbp]
    \centering
    \includegraphics[width=1.0\textwidth]{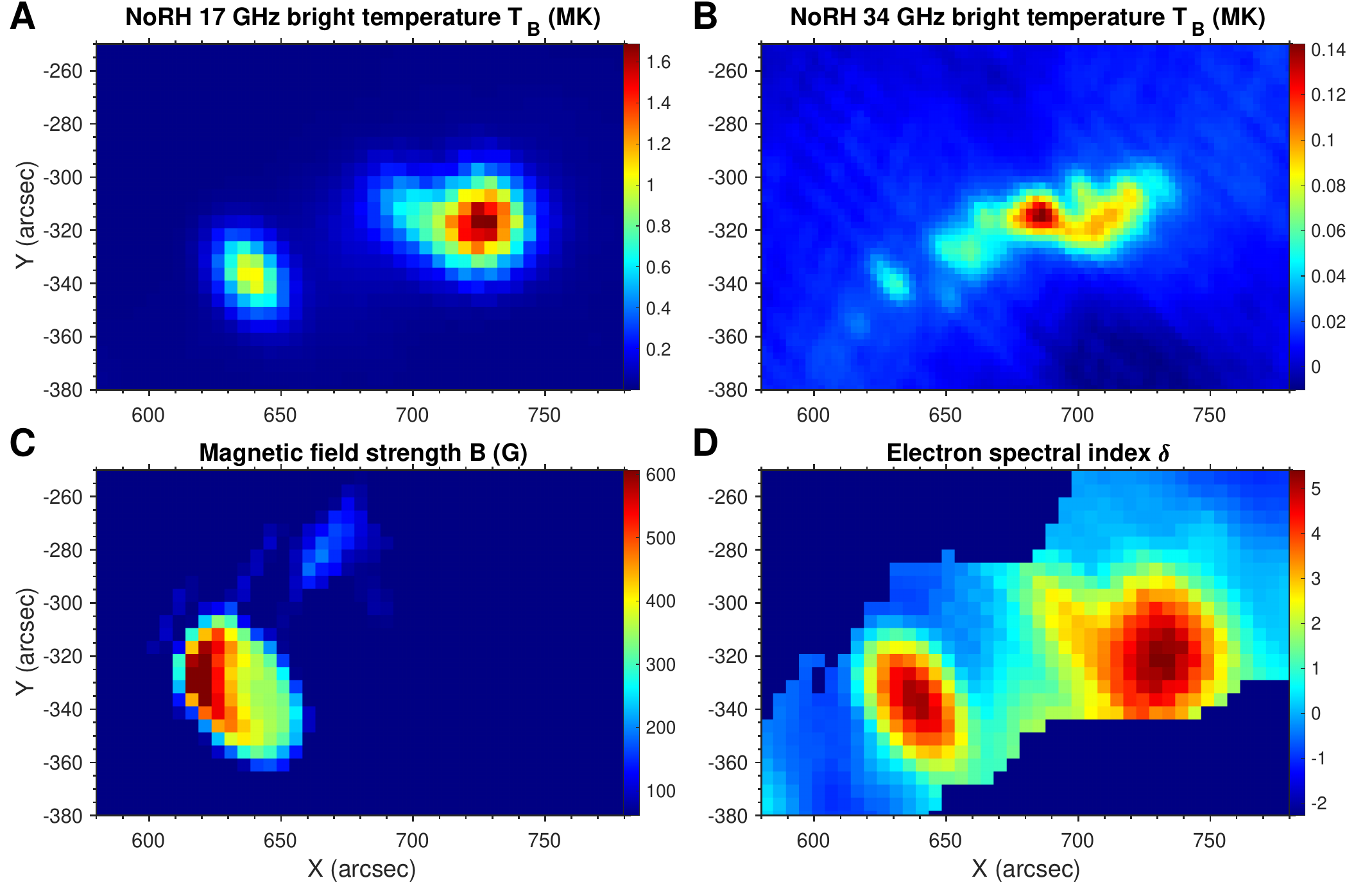}
    \caption{\textbf{Spectroscopic analysis of the microwave emission.} 
(\textbf{A}) NoRH 17 GHz bright temperature.
(\textbf{B}) NoRH 34 GHz bright temperature.
(\textbf{C}) Magnetic field strength. 
(\textbf{D}) Electron spectral index.}
\label{fig:NoRH}
\end{figure}

\begin{figure}[htbp]
    \centering
    \includegraphics[width=1.0\textwidth]{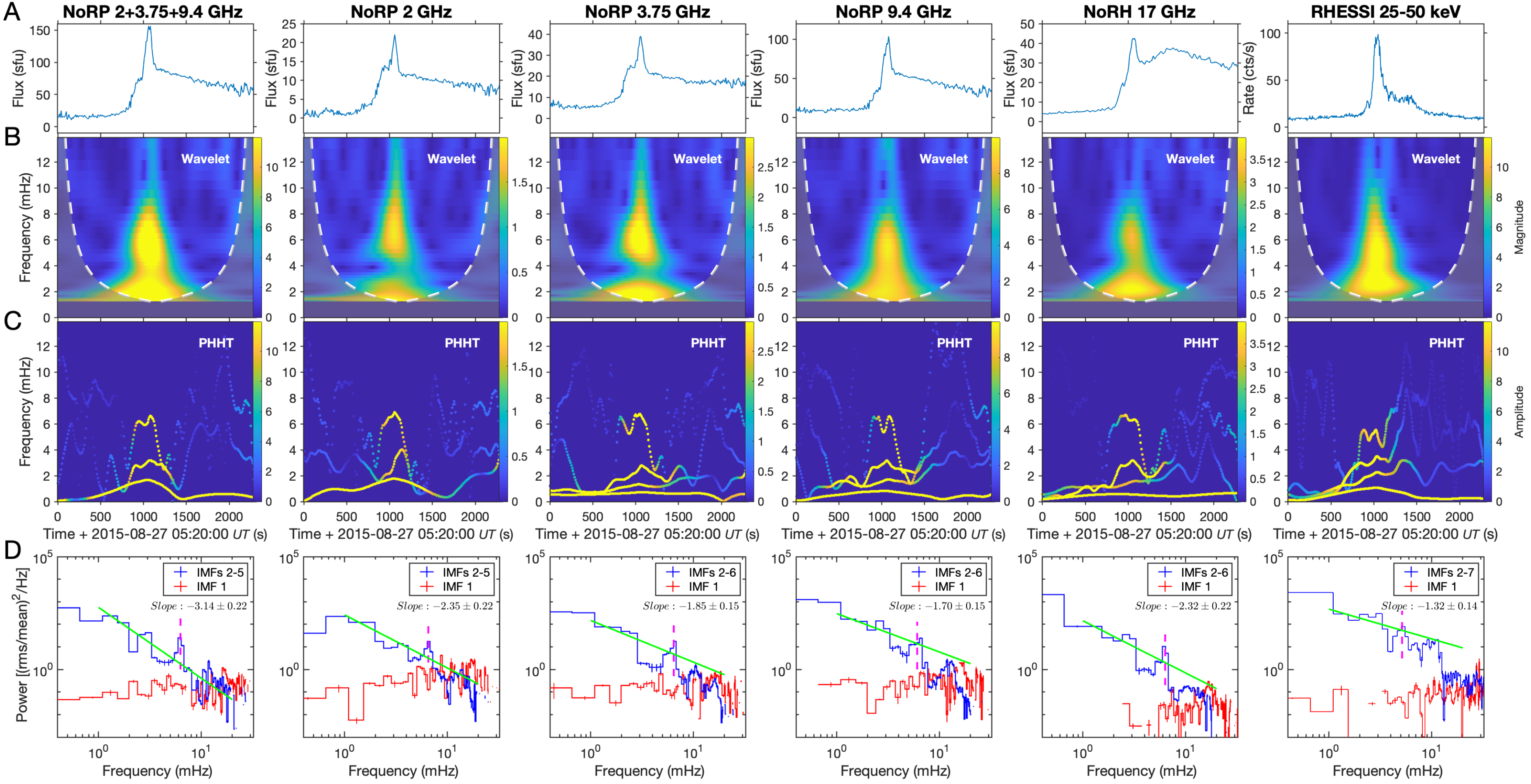}
    \caption{\textbf{Time-frequency analysis of microwave and HXR light curves.} 
	(\textbf{A}) The microwave (2, 3.75, 9.4, and 17 GHz) and HXR (25--50 keV) light curves.
	(\textbf{B}) Wavelet spectra of the light curves in (A). The white dashed line denotes the cone of influence, indicating regions where edge effects are significant. All spectra show a clear enhancement at $\sim$6 mHz.
	(\textbf{C}) Time-frequency spectra from the PHHT analysis of the light curves. All PHHT spectra also reveal the $\sim$6 mHz QPOs. To highlight these QPOs, intrinsic mode functions preceding them are not shown.
	(\textbf{D}) Marginal spectra derived from the PHHT time-frequency spectra in (C). Red lines represent the spectra of the first intrinsic mode function; blue lines show the total spectra of subsequent intrinsic mode functions. The total marginal spectra are fitted with power-law functions (green lines); their corresponding slopes and the standard deviations are indicated. The centroid frequencies of the QPOs are marked by magenta dashed lines. Error bars indicate standard deviations.}
    \label{fig:QPPwpp}
\end{figure}

\begin{figure}[htbp]
    \centering
    \includegraphics[width=1.0\textwidth]{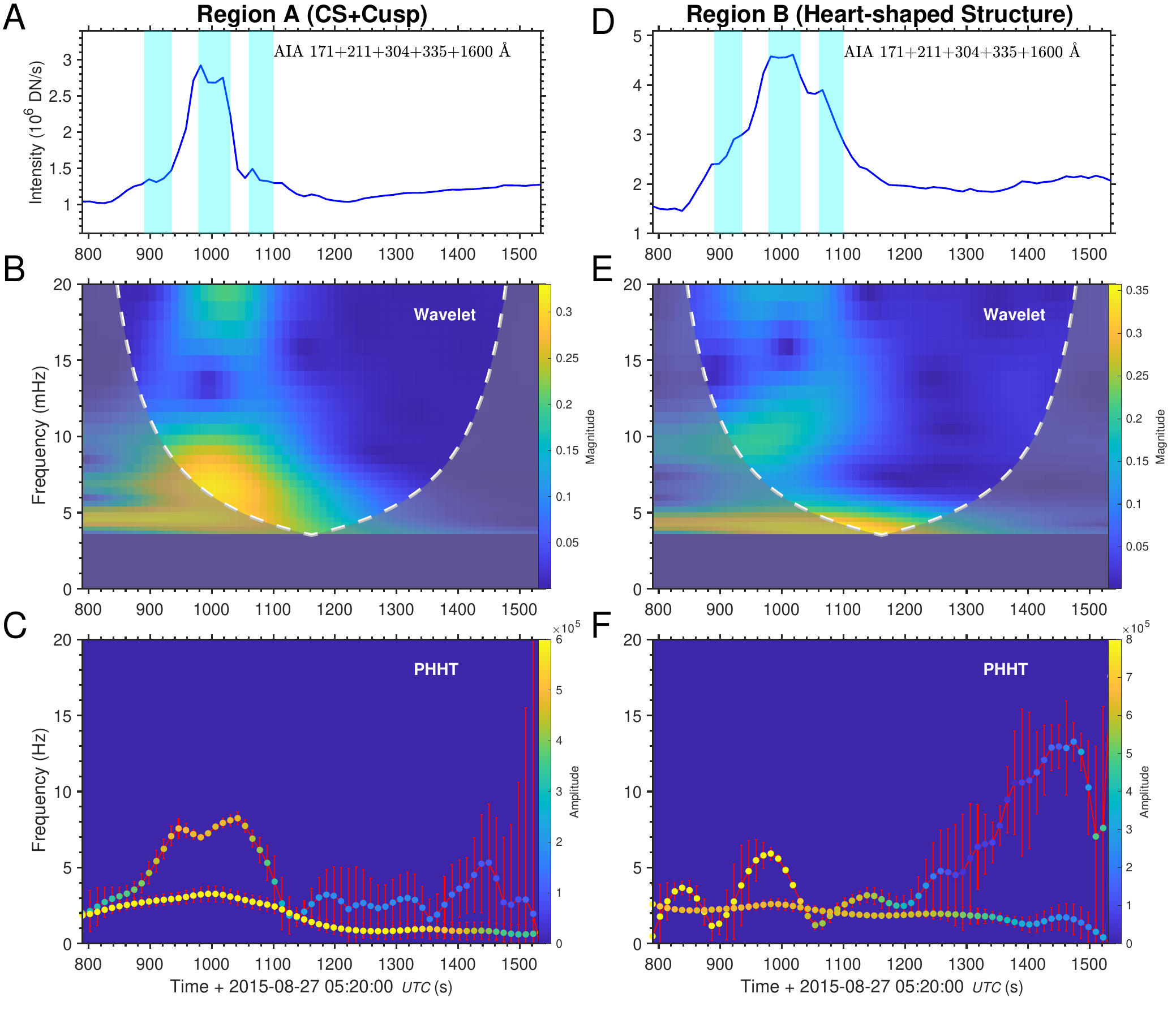}
    \caption{\textbf{Time-frequency analysis of ultraviolet light curves.} 
	(\textbf{A} and \textbf{D}) The ultraviolet (171, 211, 304, 335, and 1600 \AA) light curves from Regions A and B.
	(\textbf{B} and \textbf{E}) Wavelet spectra of the light curves. The spectrum of Region A show a clear enhancement at $\sim$7 mHz.
	(\textbf{C} and \textbf{F}) Time-frequency spectra from the PHHT analysis of the light curves. The colored curves represent averaged PHHT spectra, with red error bars indicating the standard deviations. The first intrinsic mode functions are not shown in the PHHT spectra. Cyan bars mark the three time intervals used for HXR imaging analysis. The PHHT spectrum of Region A exhibits a QPO at $\sim$7 mHz.}
    \label{fig:UVtf}
\end{figure}

\begin{figure}[htbp]
    \centering
    \includegraphics[width=1.0\textwidth]{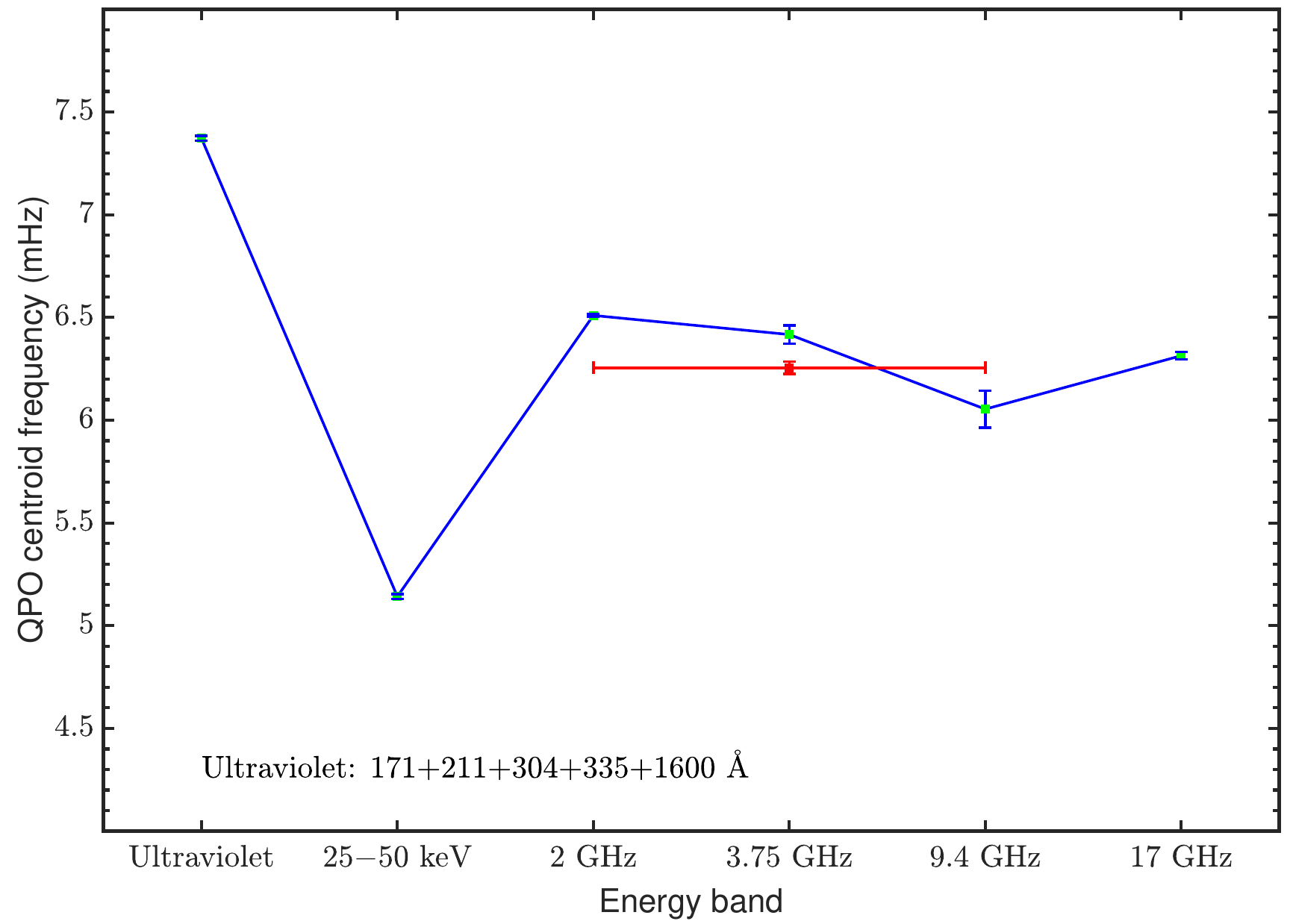}
    \caption{\textbf{QPO centroid frequency versus emission energy.} The red data point represents the QPO frequency measured in the combined microwave band (2+3.75+9.4 GHz). The error bar indicates the $1\sigma$ uncertainty in the frequency measurement.}
    \label{fig:QPPfre}
\end{figure}

\begin{figure}[htbp]
    \centering
    \includegraphics[width=1.0\textwidth]{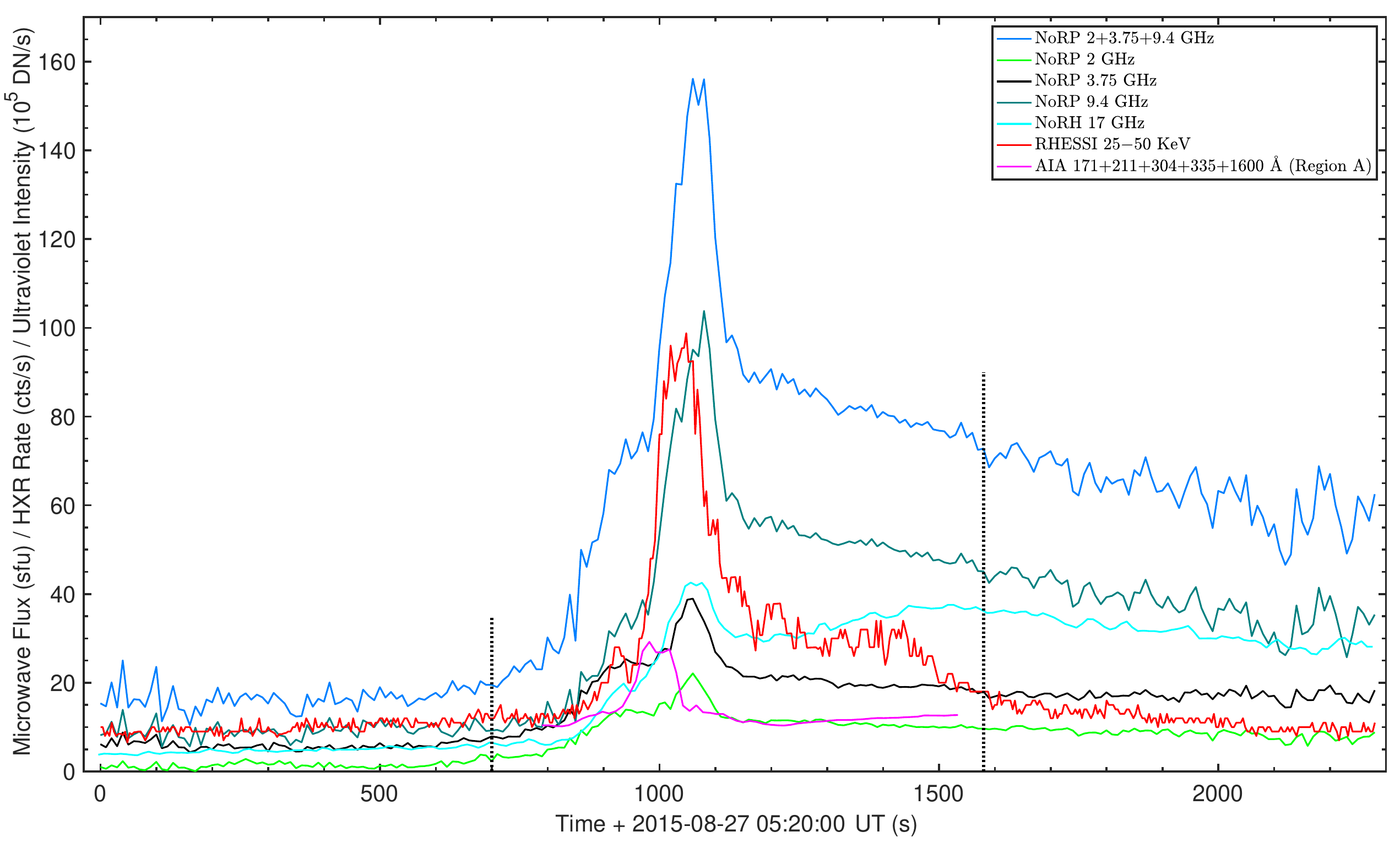}
    \caption{\textbf{Temporal ranges used in cross-correlation analysis of light curves.} The HXR, microwave, and ultraviolet light curves of the flare are shown in colors. The two black vertical lines denote the minimum range selected for cross-correlation of the microwave and HXR light curves. We systematically varied this temporal range to estimate the uncertainty in the derived time lag. While the full AIA light curve is used to perform cross-correlation with the simultaneous microwave light curve, with its lag uncertainty derived by fitting the cross-correlation peak.}
    \label{fig:LCcorrelate}
\end{figure}

\begin{figure}[htbp]
    \centering
    \includegraphics[width=1.0\textwidth]{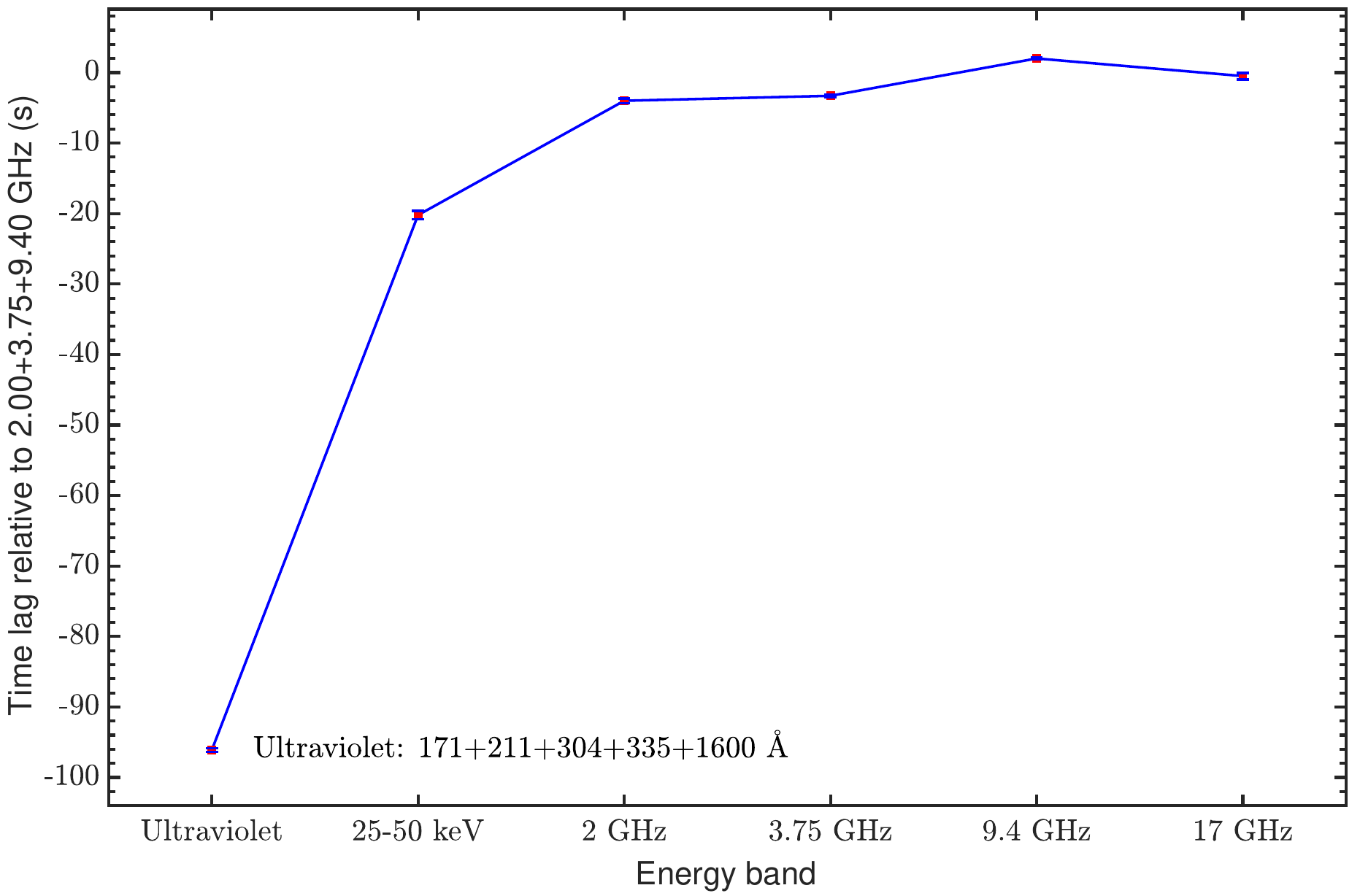}
    \caption{\textbf{Temporal lags between light curves.} The lags are derived from cross-correlation analysis using the combined microwave light curve as reference. Uncertainties are derived at the $1\sigma$ confidence level.}
    \label{fig:lags2eng}
\end{figure}

\begin{figure}[htbp]
    \centering
    \includegraphics[width=0.8\textwidth]{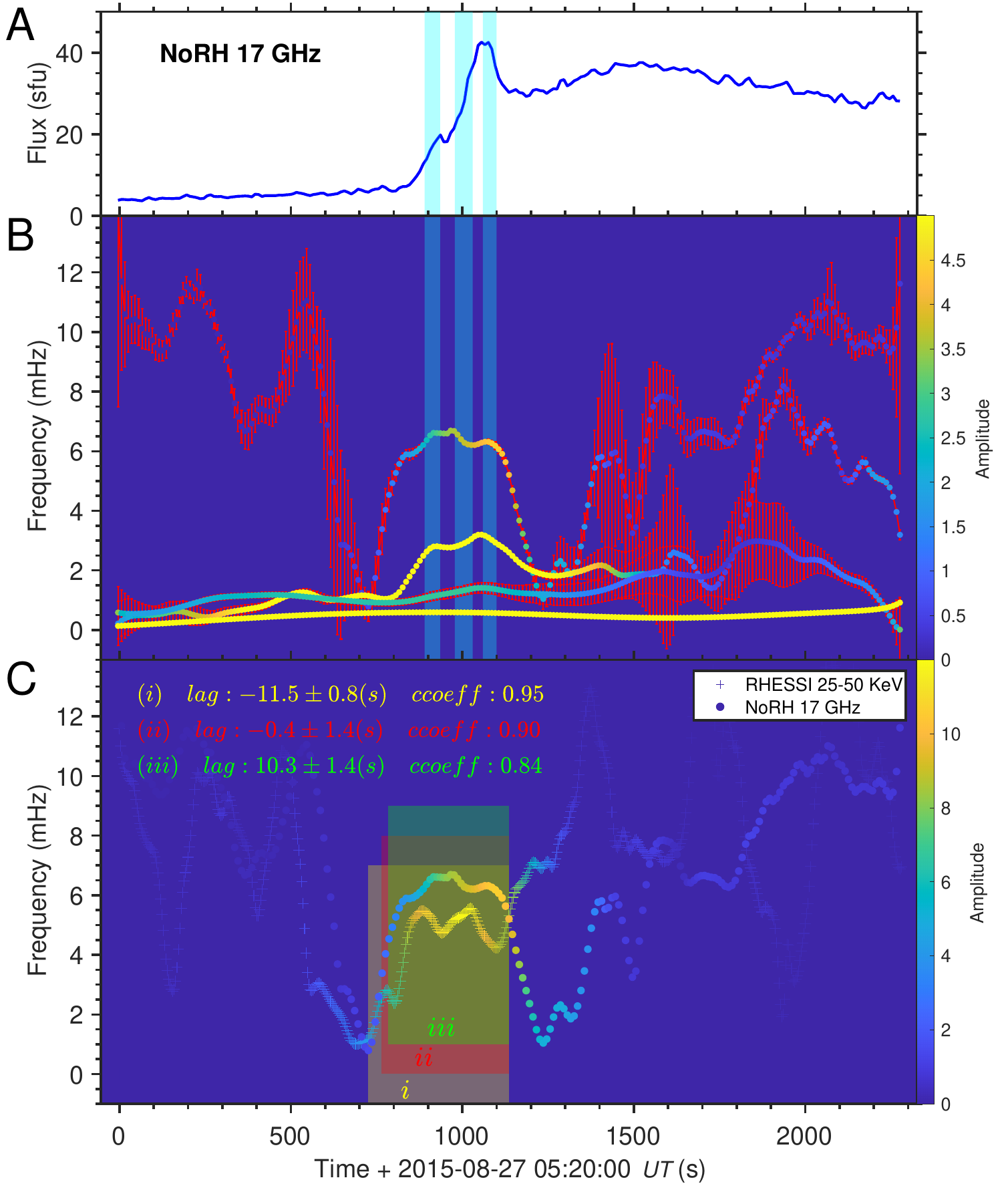}
    \caption{{\bf Temporal lag between NoRH 17 GHz QPO and RHESSI HXR QPO.} 
    (\textbf{A} and \textbf{B})  NoRH 17 GHz light curve and its PHHT spectrum. The colored curves represent averaged PHHT spectrum with red error bars indicating 1$\sigma$ standard deviations. For clarity, intrinsic mode functions preceding that containing the QPO are not shown in all PHHT spectra. Cyan bars mark the three time intervals used for HXR imaging analysis.
    (\textbf{C}) Time lag between the microwave and HXR QPOs. The HXR QPO spectrum is represented by plus sign; the microwave QPO spectrum by dots. The yellow, red, and green bars delineate three time ranges used for time-lag calculation.}
    \label{fig:QPP17G2hsi}
\end{figure}

\begin{figure}[htbp]
    \centering
    \includegraphics[width=1.0\textwidth]{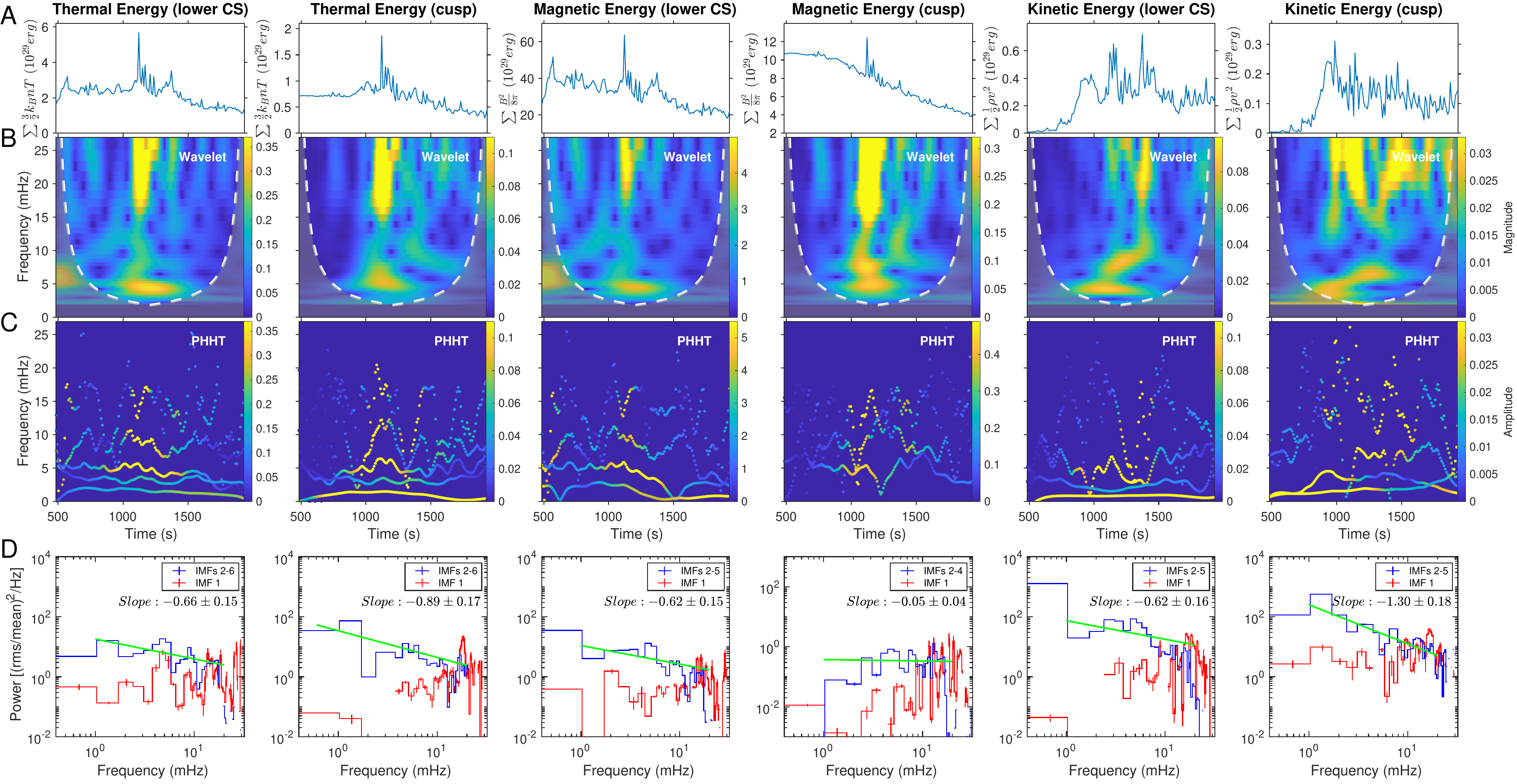}
    \caption{\textbf{Time-frequency analysis of the energy dynamics.} 
	(\textbf{A}) Temporal evolution of the thermal, magnetic, and kinetic energies integrated over the lower current sheet and cusp regions. 
	(\textbf{B}) Wavelet spectra of the energy profiles in (A).
	(\textbf{C}) Time-frequency spectra obtained with PHHT analysis of the energy profiles. For clarity, the spectra corresponding to the first intrinsic mode functions are omitted.
	(\textbf{D}) Marginal spectra derived from the PHHT time-frequency spectra in (C). The red line represents the spectrum of the first intrinsic mode function; blue lines show the total spectrum of the subsequent intrinsic mode functions. Each total spectrum is fitted with a power-law function (green line), with the resulting slope and its standard deviations indicated.}
    \label{fig:SimulEngwpp_withbkg}
\end{figure}

\begin{figure}[htbp]
    \centering
    \includegraphics[width=1.0\textwidth]{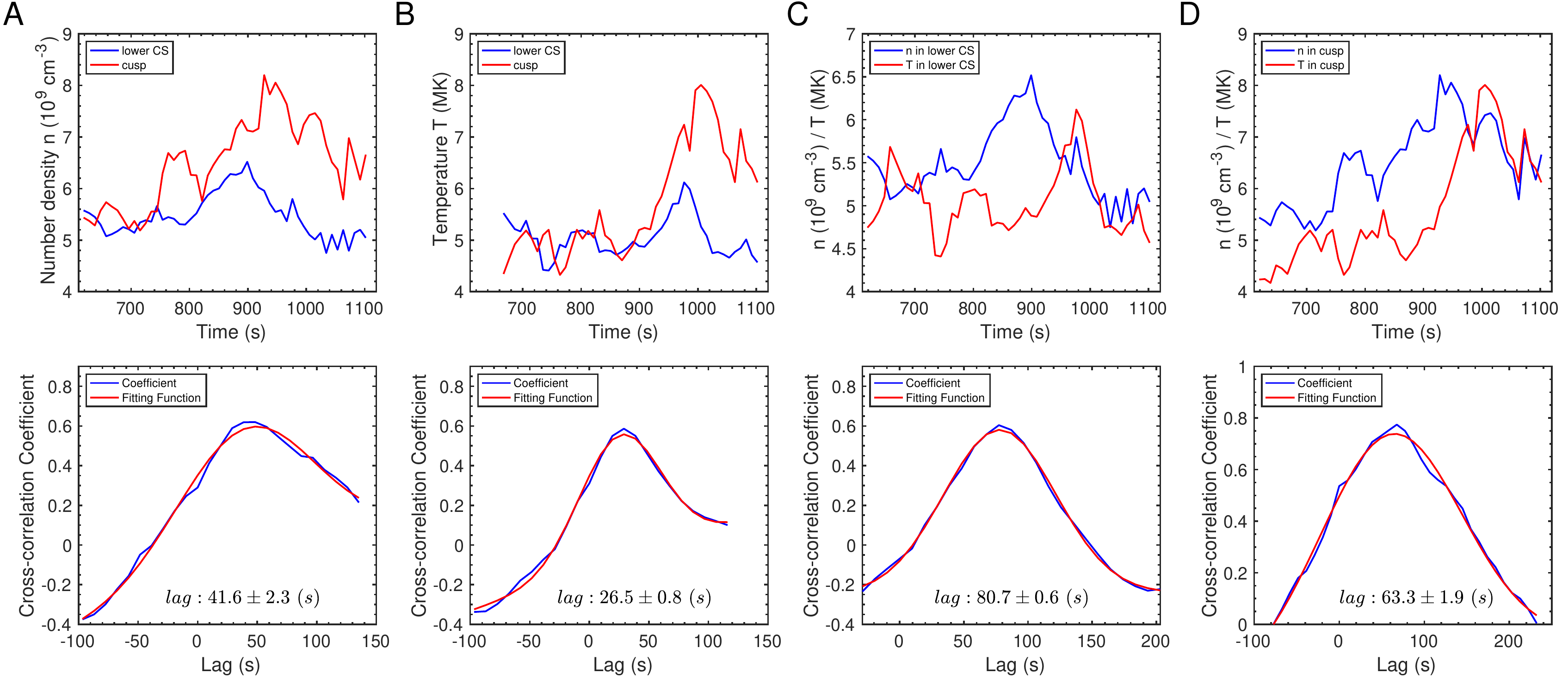}
    \caption{\textbf{Temporal lags for plasma parameters.}
	(\textbf{A}) Between number densities in the lower current sheet and cusp.
	(\textbf{B}) Between temperatures in the lower current sheet and cusp.
	(\textbf{C}) Between number density and temperature within the lower current sheet.
	(\textbf{D}) Between number density and temperature within the cusp.}
    \label{fig:Simulags}
\end{figure}

\newpage

\begin{table}
\centering
\small
\caption{\textbf{Centroid positions of X-ray and ultraviolet sources and their spatial separations.}}
\label{tab:centroids}
\begin{tabular}{@{}cccc@{}}
\hline
\textbf{Source and Time} & \textbf{Centroid Position} & \textbf{Standard Deviation} & \textbf{Distance to 1st HXR} \\
& \textbf{[arcsec]} & \textbf{[arcsec]} & \textbf{[Mm]} \\
\hline
1st HXR (34:50\text{$-$}35:45 UT) & [725.74, $-$319.93] & [2.05, 1.03] & \text{$-$} \\
1st HXR (36:18\text{$-$}37:10 UT) & [721.58, $-$320.24] & [0.91, 0.64] & \text{$-$} \\
1st HXR (37:40\text{$-$}38:20 UT) & [722.44, $-$322.11] & [1.44, 1.88] & \text{$-$} \\
\hline
Soft X-ray (34:50\text{$-$}35:45 UT) & [711.10, $-$316.46] & [0.54, 0.29] & $13.4 \pm 1.9$ \\
Soft X-ray (36:18\text{$-$}37:10 UT) & [710.16, $-$315.14] & [0.28, 0.21] & $11.1 \pm 0.8$ \\
Soft X-ray (37:40\text{$-$}38:20 UT) & [709.69, $-$314.28] & [1.29, 0.94] & $13.3 \pm 1.8$ \\
Ultraviolet cusp (36:53 UT) & [716, $-$313] & \text{$-$} & $8.1 \pm 0.7$ \\
\hline
\end{tabular}
\end{table}

\begin{table}
\centering
\scriptsize
\setlength{\tabcolsep}{2pt}
\caption{\textbf{Burst sources with QPO detections and magnetic field estimates}}
\label{QBsource0}
\begin{tabular}{@{}cccccc@{}}
\hline
\textbf{Source Name} & \textbf{Source Type} & \textbf{QPO Frequency (Hz)} & \textbf{Magnetic Field (G)} & \textbf{QPO Present Phase} & \textbf{Reference} \\
\hline
SOL2015-08-27T05:45 & Solar Flare & (5.14--7.37) $\times 10^{-3}$ & 10--25 & precursor \& rising & This study \\
SOL2017-07-13T21:56 & Solar Flare & (16.7--33.3) $\times 10^{-3}$ & 40 & -- & \cite{Kou2022} \\
MAXI J1820+070 & Black Hole Binary & 1--10 & $3 \times 10^{4} - 7 \times 10^{7}$ & -- & \cite{Bellavita2025,You2023} \\
GX 339-4 & Black Hole Binary & 0.2--8 & \makecell{$<10^{8.07}$ (accretion disc) \\ $1.5 \times 10^{4}$ (mid-infrared jet)} & -- & \cite{Belloni2005,Daly2019,Gandhi2011} \\
XTE J1550-564 & Black Hole Binary & 0.08--13, 181--281 &  $5 \times 10^{4}$ (near-infrared jet) & -- & \cite{Belloni2012,Cui1999,Li2013,Chaty2011} \\
GRS 1915+105 & Black Hole Binary & 0.002--8, 67 & $10^{3-5}$ (disc wind)  & -- & \cite{Morgan1997,Yan2013,Miller2016} \\
XTE J1859+226 & Black Hole Binary & 0.1--10 & -- & -- & \cite{Casella2004} \\
GRO J1655-40 & Black Hole Binary & 0.1--28, 274--446 & -- & -- & \cite{Motta2012,Belloni2012} \\
1ES 1927+654 & Active Galactic Nuclei  & (0.93--2.34) $\times 10^{-3}$ & 0.2 & -- & \cite{Masterson2025,Laha2025,Meyer2025}  \\
RE J1034+396 & Active Galactic Nuclei  & $2.68 \times 10^{-4}$ & -- & -- & \cite{Gierlinski2008} \\
2XMM J123103.2+110648 & Active Galactic Nuclei & $7.29 \times 10^{-5}$ & -- & -- & \cite{Lin2013}  \\
ASASSN-14li & Tidal Disruption Event & $7.65 \times 10^{-3}$ & -- & -- & \cite{Pasham2019} \\
Swift J164449.3+573451 & Tidal Disruption Event & $4.81 \times 10^{-3}$ & -- & -- & \cite{Reis2012} \\
SGR 1806-20 & Magnetar & \makecell{18, 26, 29, 92.5, 150, 626.5 \\ 720, 976, 1840, 2384} & $8 \times 10^{14}$ (dipolar) & \makecell{626.5 (precursor \& rising) \\ 1840, 2384 (transient)} & \cite{Watts2006,Strohmayer2006,Israel2005,Kouveliotou1998} \\
SGR 1900+14 & Magnetar & 28, 53.5, 84, 155.1 & $(2-8) \times 10^{14}$ (dipolar) & -- & \cite{Strohmayer2005,Kouveliotou1998} \\
SGR J1935+2154 & Magnetar & 42 & $2 \times 10^{14}$ (dipolar) & -- & \cite{Li2022,Roberts2023,Israel2016} \\
GRB 910711 & Gamma-ray Burst & 1113, 2649 & $10^{15-17}$ & precursor \& rise & \cite{Chirenti2023,Most2023} \\
GRB 200415A & Gamma-ray Burst & 2132 & -- & rise & \cite{Castro-Tirado2021} \\
GRB 230307A & Gamma-ray Burst & 909 & -- & precursor \& rising & \cite{Chen2025} \\
GRB 120323A & Gamma-ray Burst & 1258 & -- & precursor \& rising & \cite{Yang2025} \\
GRB 181222B & Gamma-ray Burst & 623 & -- & precursor \& rising & \cite{Yang2025} \\
GRB 190606A & Gamma-ray Burst & 1410 & -- & precursor \& rising & \cite{Yang2025} \\

\hline
\end{tabular}
\end{table}

\begin{table}
\centering
\scriptsize
\caption{\textbf{Sources with QPOs and associated magnetic field strengths}}
\label{QBsource1}
\begin{tabular}{@{}cccccc@{}}
\hline
\textbf{Source Name} & \textbf{Source Type} & \textbf{QPO Frequency (Hz)} & \textbf{Magnetic Field (G)} & \textbf{QPO Present Phase} & \textbf{Reference} \\
\hline
SOL2015-08-27T05:45 & Solar Flare & (5.14--7.37) $\times 10^{-3}$ & 10--25 & precursor \& rising & This study  \\
SOL2017-07-13T21:56 & Solar Flare & (16.7--33.3) $\times 10^{-3}$ & 40 & -- & \cite{Kou2022} \\
1ES 1927+654 & Active Galactic Nuclei  & (0.93--2.34) $\times 10^{-3}$ & 0.2 & -- & \cite{Masterson2025,Laha2025,Meyer2025}  \\
MAXI J1820+070 & Black Hole Binary & 2.96 & $2.4 \times 10^5 - 5 \times 10^7$& -- & \cite{You2023,Bellavita2025} \\
GX 339-4 & Black Hole Binary & 0.2--8 & $10^{6.07-8.07}$ & -- & \cite{Belloni2005,Daly2019,Gandhi2011} \\
SGR 1806-20 & Magnetar & 626.5& $8 \times 10^{12-14}$ & precursor \& rising & \cite{Watts2006,Strohmayer2006,Kouveliotou1998} \\
SGR 1806-20 & Magnetar & 2384 & $8 \times 10^{13-15}$ & transient & \cite{Strohmayer2006,Kouveliotou1998} \\
GRB 910711 & Gamma-ray Burst & 2649 & $10^{15-17}$ & precursor \& rising & \cite{Chirenti2023,Most2023} \\

\hline
\end{tabular}
\end{table}

\begin{table}
\centering
\caption{\textbf{Characteristic values for normalizing the magnetohydrodynamic equations.}}
\label{mhd}
\begin{tabular}{lcc}
\hline
Quantity & Unit & Value\\
\hline
Length & $L_0$ & $10^8$ m \\
Density & $\rho_0$ & $1.673\times 10^{-12}$ kgm$^{-3}$\\
Magnetic strength & $B_0$ & $0.003$ T\\
Pressure & $P_0$ & $7.162$ Pa\\
Temperature & $T_0$ & $2.595\times 10^8$ K\\
Velocity & $v_0$ & $2.069\times 10^3$ kms$^{-1}$\\
Time & $t_0$ & $48.33$ s \\
\hline
\end{tabular}
\end{table}


\clearpage 

\paragraph{Movie S1:}
\textbf{Differential emission measure (DEM) during the main phase of the flare.}
The DEM analysis is based on six AIA channels (94, 131, 171, 193, 211, and 335~\AA). The video presents the DEM for plasma temperatures of 1--5 MK, with HXR emission shown as red contours, illustrating the reconnection of a twisted magnetic loop.

\paragraph{Movie S2:}
\textbf{AIA~211~\AA\ imaging during the main phase of the flare.}
The video shows the process of unilateral disruption of the magnetic loop and the formation of open magnetic field lines.

\paragraph{Movie S3:}
\textbf{AIA~94~\AA\ imaging during the main phase of the flare.}
This video presents the process of plasma ejection following magnetic loop reconnection. Later in the sequence, the magnetic field configuration hosting the flare is revealed: a thick flux rope lies above and threads through a set of parallel magnetic loops beneath it, with the reconnection site located above the rope.

\paragraph{Movie S4:}
\textbf{AIA~335~\AA\ imaging during the main phase of the flare.}
This video presents the formation and evolution of the current sheet and the cusp region during the reconnection process.

\paragraph{Movies S5 and S6:}
\textbf{AIA~304~and 171~\AA\ imaging during the main phase of the flare.}
These videos present a heart-shaped magnetic structure connected to the flaring reconnecting loop. Located above the flux rope, the heart-shaped structure preexists before the flare, becomes stretched and deformed during the flare, and recovers its original shape after the flare.

\paragraph{Movie S7:}
\textbf{AIA~131~\AA\ imaging during the main phase of the flare.}
This video presents the process of plasma ejection during the flare.

\paragraph{Movies S8 to S10:}
\textbf{AIA~131, 1600, and 1700~\AA\ imaging during the main phase of the flare.}
This videos present the heart-shaped magnetic structure, current sheet, and the cusp.

\paragraph{Movies S11 and S12:}
\textbf{AIA 211 and 171~\AA\ imaging during the pre‑flare phase.}
These videos present a relatively weak UV eruption approximately 35 minutes before the flare, accompanied by the formation of the heart-shaped magnetic structure.

\paragraph{Movie S13:}
\textbf{AIA 94~\AA\ imaging during the pre‑flare phase.}
This video presents the process of two magnetic loops approaching each other and reconnecting to form the heart-shaped structure.

\paragraph{Movies S14, S15, and S16:}
\textbf{AIA 211, 171, and 94~\AA\ imaging during the post‑flare phase.}
These videos show that the flaring region is located at the intersection of the flux rope and a large magnetic loop.

\paragraph{Movie S17:}
\textbf{Flare magnetohydrodynamic simulation.}
This video presents the evolution of plasma density, temperature, and velocities in both X and Y directions during magnetic reconnection in an MHD simulation of a solar flare.

\paragraph{Movie S18:}
\textbf{Magnetic islands in current sheet in simulation.}
This video presents simulated thermal energy density images, showing the propagation of magnetic islands within the current sheet. Three dashed lines partition the region into four zones: loop, cusp, lower current sheet, and upper current sheet.



\end{document}